\documentclass{article}

\usepackage[english]{babel}

\usepackage[a4paper,top=2cm,bottom=2cm,left=3cm,right=3cm,marginparwidth=1.75cm]{geometry}

\usepackage[backend=bibtex, style=numeric, sorting=none]{biblatex}
\usepackage{amsmath}
\usepackage{graphicx}
\usepackage[colorlinks=true, allcolors=blue]{hyperref}
\usepackage{graphicx}
\usepackage{float}
\usepackage{titlesec}
\usepackage{multirow}
\usepackage{longtable}
\usepackage{lscape}
\usepackage[width=\textwidth]{caption}
\usepackage{changepage}
\usepackage{authblk}
\usepackage{hyperref}
\usepackage{multicol}
\usepackage{wrapfig}
\usepackage{csquotes}
\usepackage{caption}

\providecommand{\keywords}[1]
{
  \small	
  \textbf{\textit{Keywords---}} #1
}

\title{On Machine Learning Approaches for Protein-Ligand Binding Affinity Prediction}
\author[a,b]{Nikolai Schapin}
\author[a,b]{Carles Navarro}
\author[b]{Albert Bou}
\author[a,b,c]{Gianni De Fabritiis\thanks{corresponding author, e-mail address: g.defabritiis@acellera.com}}
\affil[a]{Acellera Labs, C/ Doctor Trueta 183, 08005 Barcelona, Spain}
\affil[b]{Computational Science Laboratory, Universitat Pompeu Fabra, PRBB, C/ Doctor Aiguader 88, 08003 Barcelona, Spain}
\affil[c]{Institució Catalana de Recerca i Estudis Avançats (ICREA), Passeig Lluís Companys 23, 08010 Barcelona, Spain}

\begin{document}
\maketitle

\begin{abstract}
Binding affinity optimization is crucial in early-stage drug discovery. While numerous machine learning methods exist for predicting ligand potency, their comparative efficacy remains unclear. This study evaluates the performance of classical tree-based models and advanced neural networks in protein-ligand binding affinity prediction. Our comprehensive benchmarking encompasses 2D models utilizing ligand-only RDKit embeddings and Large Language Model (LLM) ligand representations, as well as 3D neural networks incorporating bound protein-ligand conformations. We assess these models across multiple standard datasets, examining various predictive scenarios including classification, ranking, regression, and active learning. Results indicate that simpler models can surpass more complex ones in specific tasks, while 3D models leveraging structural information become increasingly competitive with larger training datasets containing compounds with labelled affinity data against multiple targets. Pre-trained 3D models, by incorporating protein pocket environments, demonstrate significant advantages in data-scarce scenarios for specific binding pockets. Additionally, LLM pretraining on 2D ligand data enhances complex model performance, providing versatile embeddings that outperform traditional RDKit features in computational efficiency. Finally, we show that combining 2D and 3D model strengths improves active learning outcomes beyond current state-of-the-art approaches. These findings offer valuable insights for optimizing machine learning strategies in drug discovery pipelines.
\end{abstract}


\begin{multicols}{2}
\section{Introduction}

Binding affinity is a crucial factor in the preclinical drug discovery of small-molecule therapeutics \cite{binding_aff_importance}. Traditionally assessed through experimental techniques, considerable advancements have been made in computational methods to predict protein-ligand binding affinities, including machine learned models. Nowadays, a wide range of tested model architectures exist that are employed in various predictive scenarios.

General affinity prediction models are often trained on heterogeneous binding affinity datasets, employing different input types and architectures. For instance, ligand SMILES embeddings from Large Language Models (LLMs) can be used with or without protein sequence data, as seen in ChemBoost \cite{chemboost}, DeepFusionDTA \cite{deepfusiondta}, AttentionDTA \cite{attentiondta}, and DeepDTA \cite{deepdta}. Simplified models like XGBoost \cite{xgboost1,xgboost2} or one-dimensional convolutional neural networks can then map these embeddings to affinity values. Alternatively, chemical descriptors like QM energy terms \cite{eterm1, eterm2} or physicochemical descriptors from RDKit \cite{chemdesc1, chemdesc2, rdkit} are used with simpler models such as tree-based methods, support vector machines \cite{svm}, Bayesian models, or neural networks. These can be further augmented with protein-ligand interaction fingerprints \cite{if1, if2, if3, if4, if5, if6, if7, if8, if9, if10}, capturing ligand-target interactions as 2D vectors. 3D bound conformations can also be embedded using molecular graphs and graph neural networks \cite{graphmodel1,graphmodel2,graphmodel3}, or represented through voxels, as done in KDeep \cite{kdeep} and other models \cite{voxel1, voxel2, voxel3, voxel4}. All these models can rank and prioritize compounds during virtual screening.

Among the more precise computational methods are the free energy perturbation approaches known as the alchemical transfer method (ATM) \cite{atm,atm2,atm3,atm4} or the Free Energy Perturbation (FEP) \cite{fep_schrodinger_set} method. Despite delivering state-of-the-art results in estimating binding free energies, they incur a high computational cost. This cost renders the screening of large chemical libraries impractical. To mitigate this computational burden, while still covering a broad chemical space, machine learning (ML) methods have been explored that are either trained to predict absolute or relative binding affinity. They can either use ligand-based 2D molecular descriptors with simple tree-based \cite{al1,al2,al3} or other linear models \cite{al3} or 3D bound poses \cite{kdeep,deltadelta}. 
When combined with methods like ATM, several rounds of screening and model fine-tuning may precede a round of experimental validation, thereby reducing the overall cost of virtual screening by minimizing the need for resource-intensive experimental validation.

Classifiers of binding affinity are particularly interesting for discarding non-binding molecules during virtual screening campaigns. Similar to models for general binding affinity prediction, a wide range of models can be utilized here, from those operating on LLM embeddings \cite{class1} to 2D models that utilize ligand molecular descriptors \cite{class2} or protein-ligand interaction fingerprints \cite{class3}, as well as 3D models that take the bound protein-ligand complex as input, either as a graph \cite{class4, class5} or as voxels \cite{bindscope}. A more exhaustive overview and comparison of the different embeddings and model types used for binding affinity prediction can be found in our review publication \cite{binding_aff_importance}.

Due to their flexibility, neural network models benefit greatly from supervised or unsupervised pretraining. Various approaches to perform unsupervised pretraining for chemical models are currently being explored. For example, one can derive meaningful molecular embeddings directly from textual data using a masked text learning approach on molecular SMILES or SELFIES representations, as seen in \cite{biot5+}. Alternatively, a step-wise pretraining approach may involve adding learnable input tags \cite{tagllm}, which can subsequently be reused for downstream tasks. Additionally, integrating pretrained models like protein sequence encoders \cite{pl_llm} and merging their embeddings can further enhance binding affinity predictions. Unsupervised pretraining can also be done with 3D graph-based models by predicting the atomic coordinates of bound complexes from their random conformations \cite{denoising}.

In this work, we independently train and validate multiple ML approaches, specifically: 1) graph-based 3D neural network regressors, and 2) tree-based 2D models. These models are applied in various modalities for estimating binding affinities, including general scoring and ranking of compounds, simulations of active learning cycles, and classification of binders and decoys. Each model is evaluated on public, well-known benchmarks to compare their performances. This allows us to assess how inputs, embeddings, and model types influence binding affinity prediction across different scenarios. In particular, we test on both multi-scaffold and single-scaffold ligand libraries against various targets. The different compositions of the libraries allow us to assess the models under different data types and degrees of compound sampling. The broad benchmark in this work provides deeper insights into which embeddings and model types are most effective for each use-case scenario, highlighting their advantages and shortcomings.

In order to further evaluate the usefulness of pretraining neural network models through supervised or unsupervised pretraining, we 1) compare performances of 3D graph models during active learning with and without supervised pretraining and 2) evaluate the performance of molecular embeddings generated by pretrained LLM models for binding affinity prediction. We compare the latter against classical RDKit embeddings and evaluate their coupling with different 2D ML model types.

\section{Methods}
\subsection{Datasets}\label{sec:datasets}
\subsubsection{Train Datasets}\label{sec:train_datasets}

For the general ranking experiments, models were trained on a general dataset comprised of protein-ligand complexes sourced from several publicly available datasets, including the 2020 version of the PDBbind dataset \cite{pdbbind}. The refined version of this dataset was used, which was prepared according to the steps described in their publication. 

This refined version was further expanded with complexes bearing IC50 affinity labels taken from the general PDBbind2020 set, and filtered using the same preparation steps as the refined set. Additionally, complexes from the BindingDB dataset \cite{bindingdb} were included, encompassing both the general dataset and docked congeneric series, as well as the complexes in the BindingDB validation sets. The complexes from this last group were docked using Acedock \cite{skeledock}, a tool available at \hyperlink{https://open.playmolecule.org}{Playmolecule.org}. For each ligand, 100 docking poses were generated under a restrained setting with pharmacophoric rescoring. This latter method attempts to align the ligand against a reference ligand, optimizing the overlap of pharmacophoric features on both structures and modifying the docking score based on the pharmacophoric overlap achieved. A known crystal pose of a ligand for each target served as the reference. The best docked poses were selected based on the docking score, which was amended by the pharmacophoric rescoring function.

All the complexes underwent a consistent data preparation routine that included the removal of duplicate ligand-protein complexes, exclusion of complexes where the ligand displayed multiple conformations within the binding pocket, assignment of correct protonation to the ligands and protein targets, and elimination of complexes with non-drug-like ligands based on violations of Lipinski's Rule of Five. These preparations were carried out using Aceprep, an in-house library of ligand preparation tools used in drug discovery, available through \hyperlink{https://open.playmolecule.org}{Playmolecule.org}. Following these steps, the refined dataset comprised 17,473 complexes, including 11,044 from the BindingDB dataset and 6,429 from PDBBind2020.

\begin{table}[H]
    \centering
    \resizebox{1.0\columnwidth}{!}{%
    \begin{tabular}{l|r}
        Dataset & Number of complexes  \\\hline
        Full Dataset & 17473 \\
        BindingDB & 11044 \\
        PDBBind2020 & 6429 \\
    \end{tabular}%
    }
    \captionsetup{width=1.0\columnwidth}
    \caption{\label{tab:datasizes}Size of the full training dataset and its subsets.}
\end{table}

\subsubsection{Test Datasets}\label{sec:test_datasets}

Various external datasets were employed for extensive benchmarking of different model architectures, including the PDBBind16 core set, and the JACCS \cite{fep_schrodinger_set}, OpenForceField \cite{openff}, and Merck FEP benchmark test sets \cite{merck}. The core PDBBind16 set, derived from the refined version of the PDBBind16 dataset, is prepared through clustering of the protein targets with a 90\% similarity threshold. From the largest clusters, complexes were chosen to represent those with the highest and lowest reported binding affinities, as well as additional randomly selected data points with evenly spaced binding affinities within their respective clusters. This gives a total size of 281 protein-ligand complexes.

Ligand series from the Merck \cite{merck}, OpenForceField \cite{openff}, and JACCS \cite{fep_schrodinger_set} datasets were also employed as test sets to evaluate the models. These series simulate virtual screening campaign scenarios and consist of congeneric series for different targets, where all ligands bind to the same site and share a common core scaffold with variations in the R-groups. A total of 17 protein targets are represented across the four benchmark sets, with varying numbers of ligands in each series (details provided in the Supplementary Information (SI)). Since the labels of the Merck and JACCS datasets, as well as some from the OpenForceField, were expressed as $\Delta$G values, they were converted to $K_d$ values using the following formula: $\Delta G = RT\ln(K_d)$\label{eq:delta_g_conversion}, with R the universal gas constant and T=298K the temperature. Also, because the tested embeddings from the BERT LLM model (described in section \ref{sec:schrodinger_merck_results}) were pretrained on only non-charged molecules, charged molecules from the JACCS test sets for targets Tyk2, CDK2, Jnk1 and p38 were excluded to provide a fair comparison between all models.

Additionally, to evaluate the performance of the ML models on a large congeneric ligand series dataset that mimics a real-world scenario of active learning, we utilized the dataset from \cite{al1}. This dataset comprises a congeneric series of 10,000 ligands, each with binding affinities against the Tyk2 target, computed using an FEP-like method. All ligands were docked against the target using AceDock \cite{skeledock} in a restrained manner, using the smallest ligand from the Tyk2 JACCS congeneric series as the reference molecule. For each ligand, 20 poses were generated and rescored based on pharmacophoric overlap with the reference ligand, and the best-scored pose was selected. A small subset of ligands, which were smaller than the reference, could not be accommodated through restrained docking. These were redocked in a free docking manner, utilizing the reference ligand solely to specify the binding pocket. Binding affinities, represented as $\Delta$G values, were converted using the formula mentioned above.

\subsection{Tested Models}

We tested several model architectures, including three graph-based models and one tree-based model. Among the graph-based models, we examined the Schnet architecture \cite{schnet}, hereafter referred to as GraphNet, an equivariant graph neural network transformer \cite{torchmdnet}, named Equivariant Transformer for the remainder of this work, and the TensorNet model architecture \cite{tensornet}. 

\begin{figure}[H]
\centering
\includegraphics[width=1.0\columnwidth]{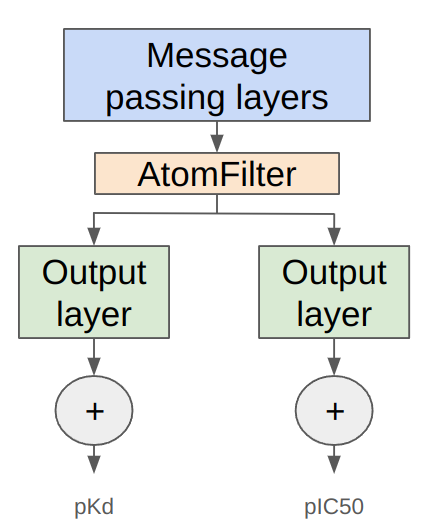}
\captionsetup{width=1.0\columnwidth}
\caption{Schematic representation of the modified output layer. The message passing layers remain unchanged, facilitating interactions among various atoms in the input structure. Learned latent embeddings of the ligand atoms are processed through the output layers, where each is reduced to a scalar value representing its individual contribution. These contributions are subsequently summed to produce the predicted binding affinity values. Separate heads can be constructed and learned in parallel for each binding affinity metric.}
\label{fig:mod_output_layers}
\end{figure}

To facilitate differentiation between ligand and binding pocket atoms, we introduced separate embeddings for ligand and protein atoms based on their atomic elements. The message passing layers were retained as per their original implementations to facilitate the learning of quantum mechanical molecular features. However, the final output head was modified, which consists of two linear layers with a non-linear activation function interposed. A flexible, multi-head architecture was enabled to allow for parallel learning of various binding affinity metrics, as shown in figure \ref{fig:mod_output_layers}. This modified head processes only ligand atoms' latent embeddings, transforms them into a single scalar value and applies a final sum pooling operation over the learned scalar contribution values of each ligand atom.

For the tree-based model, we utilized the XGBoost architecture. The input consisted of a linear concatenated embedding vector that included three structural molecular fingerprints—Morgan, MACCS, and Avalon—and one vector of molecular chemical features, all computed using RDKit \cite{rdkit}. Unlike the graph-based models, this embedding vector solely accounted for the input ligand. The hyperparameters for all models were optimized, and details of the best hyperparameters selected for each model are discussed in SI.

\section{Experiments and Results}
\subsection{General Binding Affinity Models}\label{sec:gen_aff}

\begin{figure}[H]
\centering
     \includegraphics[width=1.0\columnwidth]{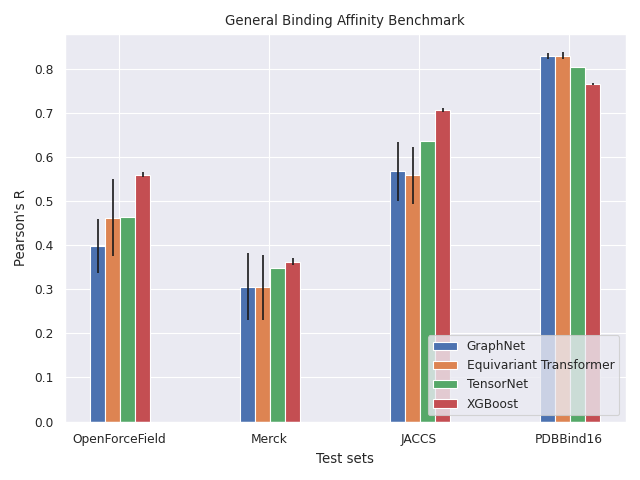}
     \captionsetup{width=1.0\columnwidth}
      \caption{Benchmark of general affinity models: Various model architectures were trained using collected binding affinity data and evaluated across different prepared test sets. Eight replicas were trained for each model, and the error bars in the figure represent the standard deviation of the predicted metric across these replicas. Due to longer training times, only one replica was trained for the TensorNet architecture.}
       \label{fig:bench_general_pearson}
\end{figure}

First, we evaluated the efficacy of different model architectures in learning general affinity information from the provided data. The models were trained using prepared datasets from PDBBind2020 and BindingDB, and subsequently tested against various external test sets. The results are presented in figure \ref{fig:bench_general_pearson}.

From the results, we observe that all three graph-based models achieve comparable performance across the different benchmark sets, considering the standard deviation across multiple model replicas. Additionally, the high performance of the tree-based model is noteworthy, even though it does not explicitly account for protein information. Given that the benchmark sets include ligands bound to various protein targets and binding pockets, this suggests that high performance may not solely stem from learned interactions between the ligand and the protein target, but could also be influenced by spurious correlations present in the training data. This observation aligns with previous research addressing this issue \cite{bias_aff_data1, bias_aff_data2}.

To further investigate this issue, we continued training both the graph-based equivariant transformer and the tree-based models on the same training data used in the previous experiment, but augmenting it with inactive decoy ligands. Given the ease of generating decoys that are structurally or physicochemically distinct from true binders \cite{bias_dude_1, bias_dude_3, bias_aff_data2}, we employed cross-docking to create the decoy subset. Initially, we clustered the proteins using BLAST \cite{blast} with a cutoff of 0.7 and the ligands using Taylor-Butina clustering \cite{taylor_butina} based on their Morgan fingerprints \cite{morgan_fp} with a radius of 3 and size of 2048, setting a Tanimoto similarity \cite{tanimoto} cutoff of 0.9. Each true binder was then paired with four decoy ligands, selected randomly from true binders that bind to protein targets in different protein clusters. We ensured that the four decoys for each binder did not come from the same ligand cluster, thereby maintaining both true binders and decoys within the same structural and physicochemical space while ensuring sufficient heterogeneity of decoys per target. This design makes it challenging for models to distinguish between the two groups based solely on ligand information.

Our rationale for the cross-docking approach is grounded in the observation that many ligands binding to structurally similar protein targets are unlikely to exhibit equally effective binding to structurally diverse targets. We retained the same benchmark test sets as in the previous experiment to specifically assess the impact of data augmentation on general binding affinity predictions.

\begin{figure}[H]
\centering
     \includegraphics[width=1.0\columnwidth]{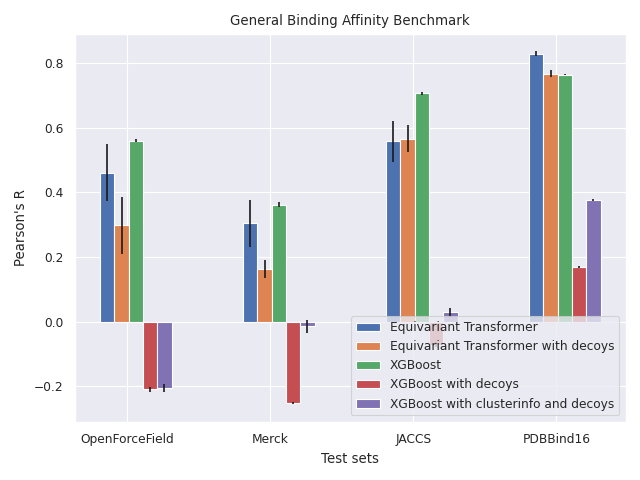}
     \captionsetup{width=1.0\columnwidth}
      \caption{Benchmark of general affinity models with decoys: The equivariant transformer and ligand-based XGBoost models were tested using the addition of decoy molecules that are structurally and physicochemically similar to the active compounds. Three replicas were trained for the models including decoys. Error bars represent the standard deviation of the predicted metric across the model replicas' predictions.}
       \label{fig:bench_decoy}
\end{figure}

From the results presented in figure \ref{fig:bench_decoy}, we observe that the graph-based model maintains comparable performance to that of the model trained on the non-augmented dataset for the JACCS and PDBBind16 core test sets. While, a performance drop is noted for the other test sets this is less pronounced than in the tree-based model, which shows large decline across all test sets. This likely stems from the fact that ligand-based descriptors fail to provide adequate information to discern differences in binding affinity among very similar ligands. This further substantiates the notion that models relying solely on ligand features tend to learn from biases in the training data, rather than from information pertinent to the binding affinity prediction task, when the training data include heterogeneous binding affinity data across multiple protein targets. Similar data augmentation techniques have also shown \cite{decoy_use} to mitigate these ligand biases and enhance the generalizability of 3D models across various industrial protein target datasets.

It is also noteworthy that the addition of simple protein information to the tree-based model, only marginally improves its performance when trained on the augmented data. This enhancement was attempted by clustering the protein targets in both the training and test sets using BLAST \cite{blast}, this time with a cutoff of 0.9, and incorporating the cluster ID as a one-hot encoded vector into the model. A higher cutoff was employed to ensure that each cluster contained the same or very similar proteins, allowing the cluster ID to serve as a proxy for protein identity. The minimal improvement observed upon adding this information suggests that protein data must embed more detailed structural and physicochemical characteristics of the binding pocket to be effectively utilized.

\subsection{Influence of Hydrogen and Waters on 3D Model's Performance}

Waters have been proven \cite{crystal_water_importance1,crystal_water_importance2,crystal_water_importance3,crystal_water_importance4,crystal_water_importance5} to play a crucial role in the binding affinity process between a ligand and its target, primarily due to their influence on the thermodynamics of binding through desolvation effects and their ability to facilitate various interactions, such as hydrogen bonds. Consequently, incorporating both crystal waters and hydrogen atoms is expected to enhance the performance of 3D models. To test this hypothesis, we evaluated the performance of the 3D GraphNet model, trained both with and without the inclusion of crystal waters and hydrogens, across the different benchmark test sets described in section \ref{sec:test_datasets}.

\begin{figure}[H]
\centering
     \includegraphics[width=1.0\columnwidth]{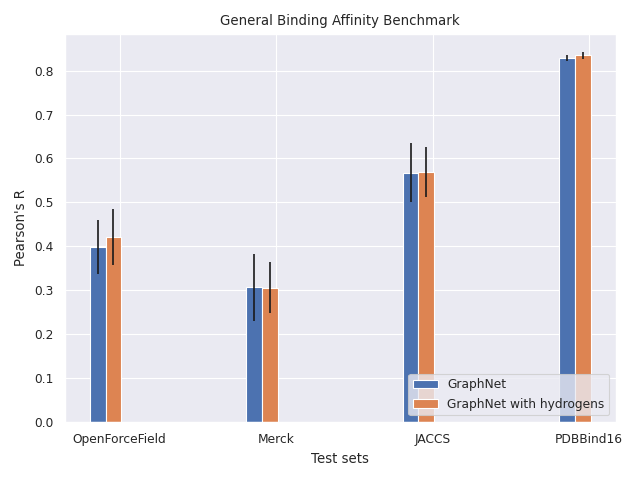}
     \captionsetup{width=1.0\columnwidth}
      \caption{Comparison of the 3D GraphNet model performance when trained with and without crystal waters and hydrogen atoms. The positions of both elements were sourced from the provided crystal structures and the prepared test sets. 8 model replicas were trained for each model and error bars represent standard deviation on the computed metric across model replicas.}
       \label{fig:bench_water}
\end{figure}

From figure \ref{fig:bench_water}, we observe that the inclusion of crystal waters and hydrogen atoms improves the results of the 3D model to only a minor extent. This modest improvement may be attributed to the often arbitrary placement of hydrogens and crystal waters during the preparation of 3D poses from crystal structures, or their failure to be readjusted during docking. Since the 3D model is sensitive to precise atomic positions, even when introducing small noise to enhance robustness, it can prevent the model from accurately learning the relationship between crystal waters and binding affinity. To more effectively utilize information from waters and explicitly added hydrogens, their positions should be recalibrated, which can be achieved through energy minimization using quantum mechanics (QM) or molecular dynamics (MD).

\subsection{Active Learning}

It is also possible to use machine learning (ML) models within active learning cycles.  To evaluate the performance of various ML architectures in an active learning scenario, we conducted several experiments using different ligand series for multiple targets. For each experiment, we trained or fine-tuned the ML model using several ligands from each series, then tested the model's performance on the remaining ligands. This process was repeated for several rounds, successively adding the newly chosen ligands from the ligand pool to the growing training set. We tested ligand series comprising either multiple scaffolds or congeneric series with a fixed scaffold and R-group modifications.

\subsubsection{Multi-Scaffold Sets}\label{sec:multi-scaffold}

\begin{figure}[H]
\centering
     \includegraphics[width=1.0\columnwidth]{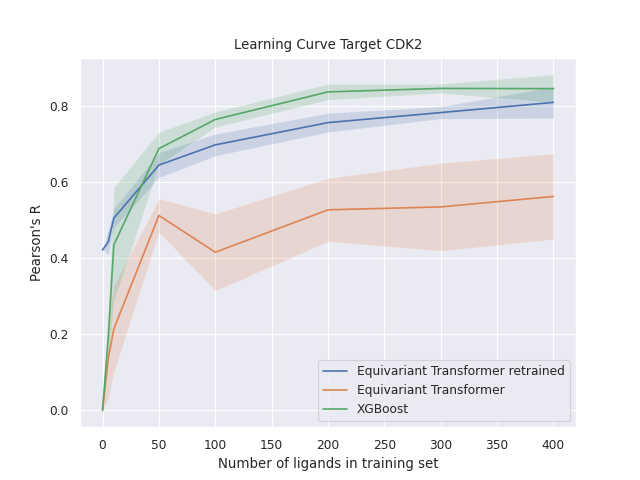}
     \captionsetup{width=1.0\columnwidth}
      \caption{Learning curve for CDK2 target: Incremental training was performed using binding affinity data for the specific target, simulating a virtual screening scenario in computational drug discovery where a library of potential ligands is screened against a designated target. This figure compares the performance of the graph-based equivariant transformer model and the tree-based model. Additionally, it contrasts a graph-based model trained solely on target-specific data with one pretrained on general binding affinity data from various other targets. While the tree-based model, which relies only on ligand information, initially lags, it quickly approaches the performance of the pretrained graph-based model as more ligand data become available for the specific target. Shaded area represents standard deviation on the computed metric across 5 model replicas.}
       \label{fig:lc_cdk2}
\end{figure}

First we generated multi-scaffold ligand series for several targets. The pool of ligands was derived by clustering the protein targets in the training set described in section \ref{sec:train_datasets} using BLAST with a threshold of 0.9. We also verified the protein target names in PDB \cite{pdb} and BindingDB to confirm that all ligands were binding to the same target within each pool. This pool was further enriched with ligands from the test sets mentioned in section \ref{sec:test_datasets} that bind to the corresponding targets. In each iteration, 50 ligands were randomly selected to be added to the training set, except for the first iteration where only 10 ligands were chosen. Each cycle included 5 training runs to account for differences in ligands selection. Figure \ref{fig:lc_cdk2} displays the learning curve obtained for the target CDK2 and learning curves for other targets can be found in SI.

The tested models included the equivariant transformer graph-based model and the tree-based model. Unlike the experiments in section \ref{sec:gen_aff}, where the models' scoring abilities were tested in a general setting across different protein targets, in this active learning setup, ligands are typically assessed for the same protein target. Therefore, models like the tree-based model, which relies only on ligand features, are suitable as they can implicitly infer the fixed binding pocket environment from the ligands' structures. Additionally, we compared a graph-based model pretrained on the general training data (as described in section \ref{sec:gen_aff}) with a non-pretrained model initialized with random weights. Both the protein targets and ligands from the active learning pools were excluded from the general training data when pretraining the model.

\begin{figure*}[!ht]
\centering
     \includegraphics[width=\textwidth,height=0.368\textheight,keepaspectratio]{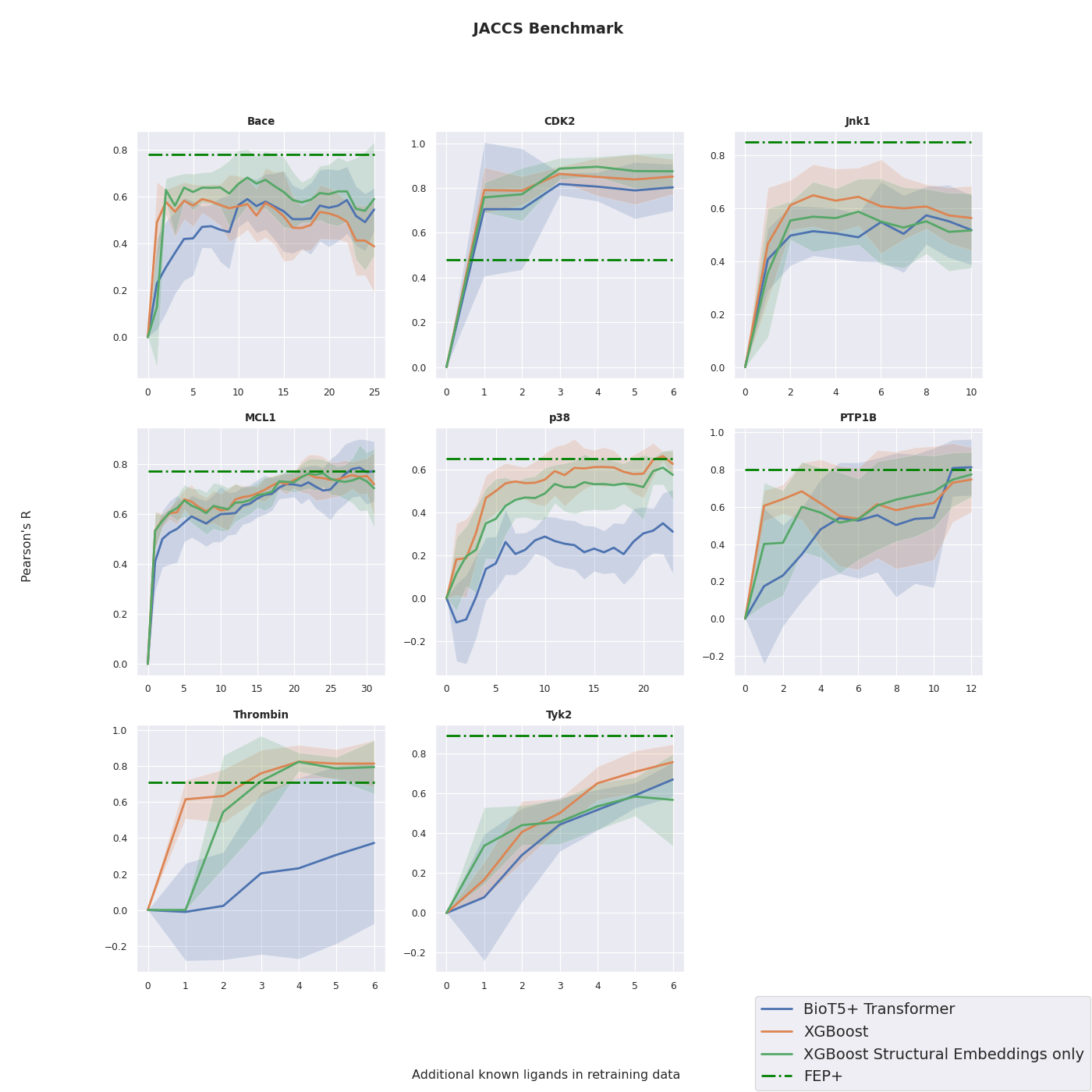}
      \caption{Learning curves for JACCS targets: Learning curves were generated for several target test sets within the JACCS FEP benchmark sets. This figure compares the performance of the XGBoost model to that of a transformer model utilizing a pretrained LLM embedding. As the LLM embedding includes only structural information of the ligand, results from an XGBoost model employing solely structural RDKit embeddings are also depicted for a direct comparison. Plotted is the Pearson's correlation against the pool of test molecules. The shaded area represents the standard deviation in the computed metric across the 25 model replicas. Included is also the performance of FEP+ on predicting absolute binding affinity as a reference.}
       \label{fig:schrodinger_bench_pears}
\end{figure*}

\begin{figure*}[!hb]
\centering
     \includegraphics[width=\textwidth,height=0.368\textheight,keepaspectratio]{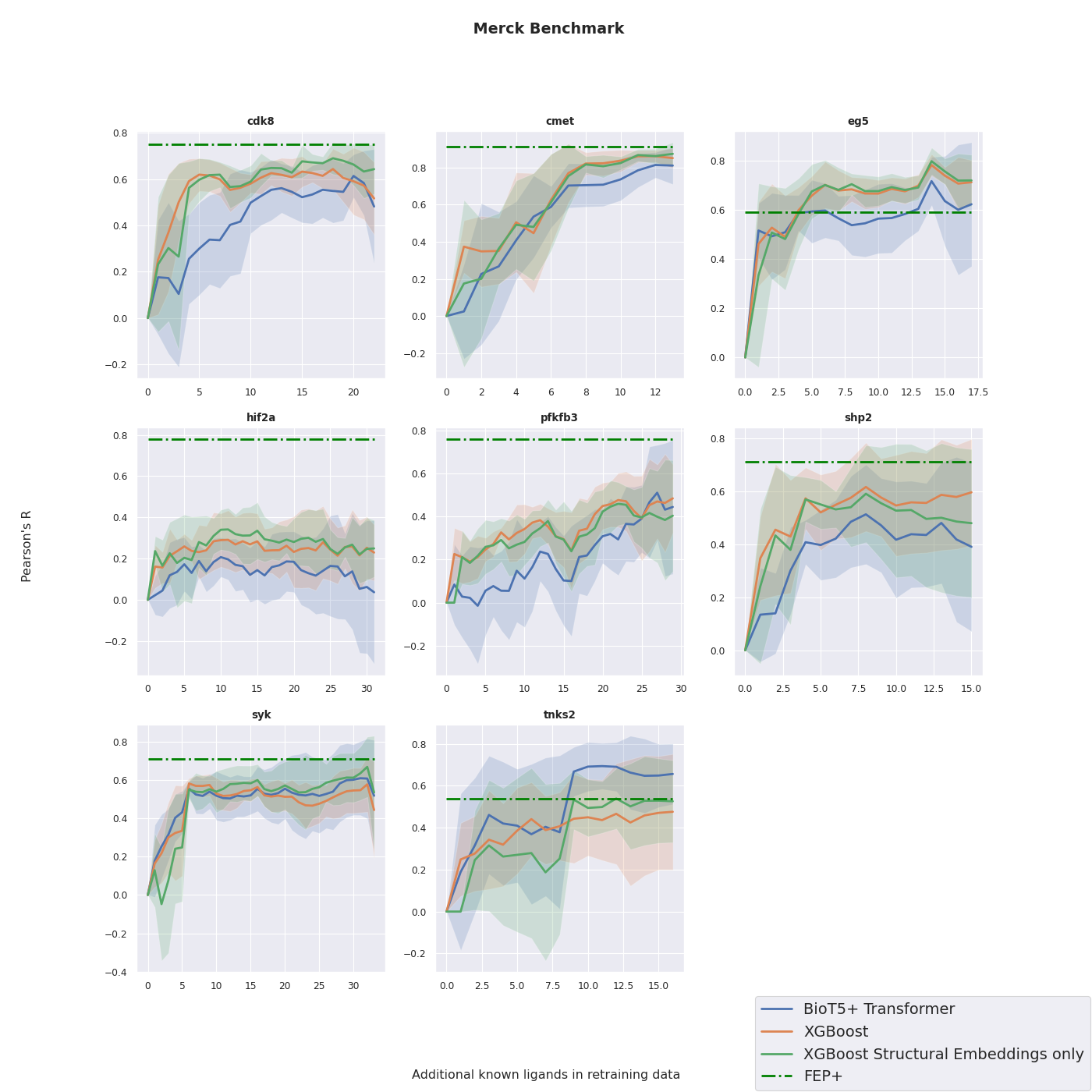}
      \caption{Learning curves for Merck targets: Learning curves were generated for multiple target test sets within the Merck FEP benchmark sets. This figure compares the performance of the XGBoost model with a transformer model that utilizes a pretrained LLM embedding, which contains only structural information of the ligand. For comparison, results for an XGBoost model employing only structural RDKit embeddings are also shown. Plotted is the Pearson's correlation against the pool of test molecules. The shaded area represents the standard deviation in the computed metric across the 25 model replicas. Included is also the performance of FEP+ on predicting absolute binding affinity as a reference.}
       \label{fig:merck_bench_pears}
\end{figure*}

From the learning curves, we first observe that performance improves for all models as more ligands are added to the training set. Notably, the pretrained graph-based model consistently outperforms a randomly initialized model across all data splits, highlighting the benefits of supervised pretraining on heterogeneous binding affinity data. This pretraining allows the model to extract general binding affinity relationships from other targets' data, providing a baseline scoring performance even without target-specific training, thereby demonstrating a degree of learned generalizability.

Conversely, the tree-based model, tailored specifically to the given target, initially shows limited scoring ability due to the sparse data available. However, as it receives more ligand data with binding affinity information, its predictive performance rapidly improves, eventually surpassing the graph-based model once 50 binding affinity data points are available.

\subsubsection{JACCS and Merck FEP Sets}\label{sec:schrodinger_merck_results}

Similar learning curves were constructed for the different ligand sets in the JACCS and Merck FEP benchmark sets. Due to the limited size of the congeneric series in these sets, one ligand was selected at a time to be included in the training set. 25 model replicas were trained during each cycle to account for random ligand selection and model randomness. Unlike previous tests, these series are congeneric, meaning all ligands share the same scaffold with variations in R-group modifications. We continued testing with the XGBoost model, which had shown strong performance in earlier evaluations.

In previous experiments, while pretraining a 3D model led to improved performance on unseen ligand series, the performance gains did not increase as rapidly as those seen with the XGBoost model. To explore whether more extensive unsupervised pretraining could enhance the performance of neural network-based models, we compared the XGBoost model to a transformer model. This transformer model utilizes ligand embeddings from a pretrained Large Language Model (LLM), specifically the BioT5+ model \cite{biot5+} trained on extensive biomedical data. These embeddings are processed by a transformer head featuring BERT-like attention mechanisms \cite{BERT}, designed to extract the relevant information from the ligand embeddings and map it to the respective binding affinities.

From the learning curves in figures \ref{fig:schrodinger_bench_pears} \& \ref{fig:merck_bench_pears}, we observe that the XGBoost model's performance rapidly improves as more ligands from each series are introduced. Additionally, we compared its performance with the FEP+ method from Schrodinger \cite{fep_schrodinger_set}, which estimates $\Delta$G using a free energy perturbation method similar to ATM. This comparison reveals that the XGBoost model achieves comparable or superior performance to the more computationally intensive FEP+ method for some target sets.

\begin{figure*}[!hb]
\centering
     \includegraphics[width=\textwidth,height=0.4\textheight,keepaspectratio]{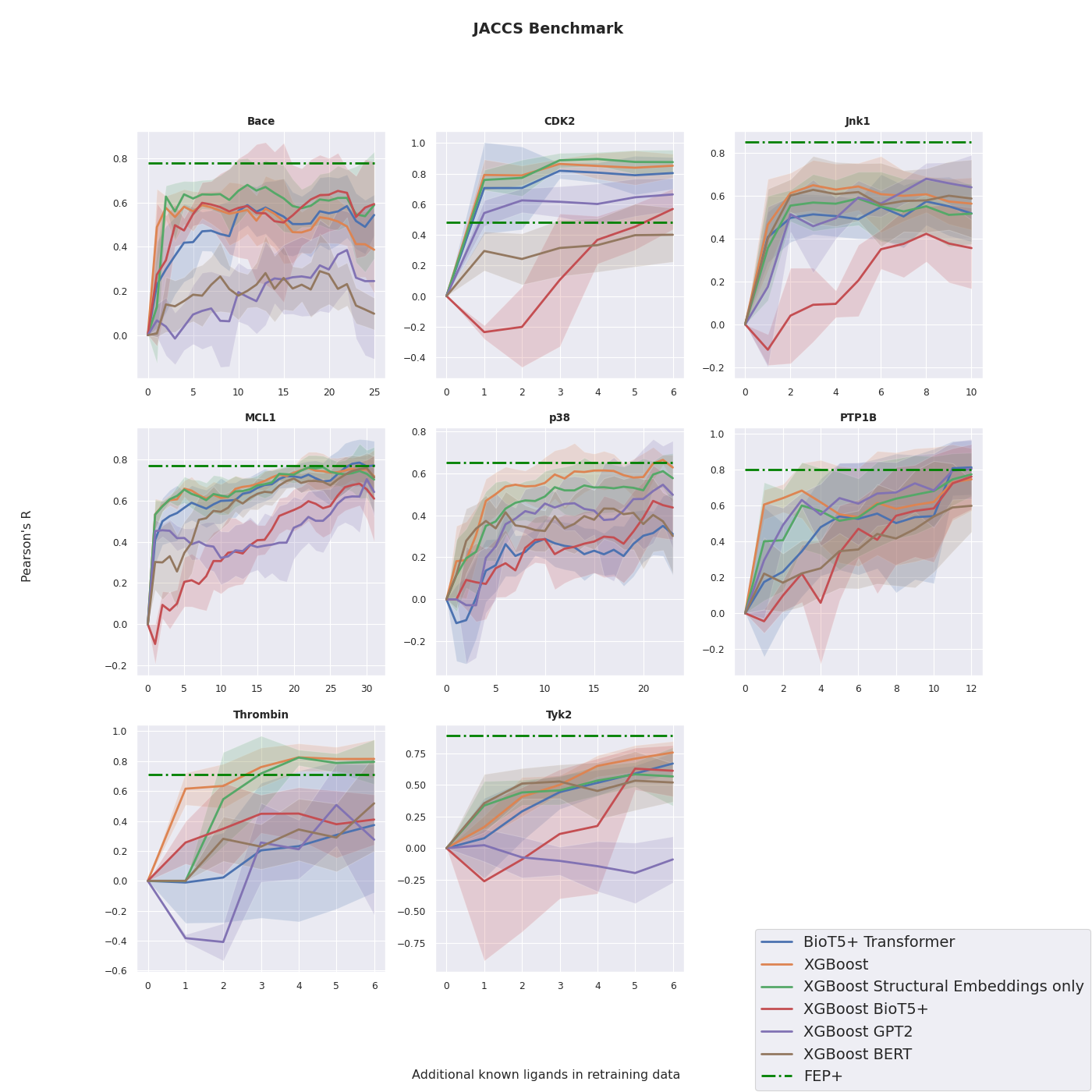}
      \caption{Additional learning curves on the JACCS benchmark sets comparing XGBoost models. This comparison features models using RDKit embeddings against those with LLM embeddings from various architectures. Given that LLM embeddings encode only structural information, the figure also includes the performance of an XGBoost model utilizing solely structural RDKit embeddings. Plotted are Pearson's correlations against the pool of test compounds. The shaded area represents the standard deviation in the computed metric across the 25 model replicas. Included is also the performance of FEP+ on predicting absolute binding affinity as a reference.}
       \label{fig:schrodinger_bench_pears_extra}
\end{figure*}

In contrast to the previous experiment involving a larger, multi-scaffold CDK2 ligand set, the JACCS and Merck test sets show a much quicker performance improvement, with Pearson’s correlation exceeding 0.6 after training with as few as five ligands for some targets. This disparity in performance acceleration can likely be attributed to the different compositions of the ligand sets. While the multi-scaffold CDK2 set included a more diverse array of ligands, the ligands in the JACCS and Merck sets are part of congeneric series, thus presenting a less varied chemical space. This restricted space allows the tree-based model to learn quicker from the data.

When testing the XGBoost model against the pretrained LLM embeddings, we observed that performance with LLM embeddings is comparable to that achieved using classic RDKit descriptors for some of the ligand series. This suggests that the LLM successfully captures useful chemical information within the embeddings, which can then be effectively utilized by attention head mechanisms for specific downstream tasks. However, poorer performance on some targets may be attributed to the small sizes of certain ligand series, as the mean absolute error differences between the models are minimal (see SI for additional metrics).

To further evaluate the BioT5+ embedding against other LLM architectures, figure \ref{fig:schrodinger_bench_pears_extra} presents the results of XGBoost models using embeddings from GPT2 \cite{gpt2} and BERT \cite{BERT} pretrained LLM models on the JACCS benchmark sets. The GPT2 model, pretrained on 400M small molecule SMILES string representations from the Enamine REAL 350/3 dataset \cite{enamine_real}, produces per-token embeddings of a dimension of 256. While the BERT model, \\\\\\ pretrained on 1.2M small molecule SMILES strings from CHEMBL \cite{chembl} using a contrastive loss, produces a single embedding vector with a dimension of 128.

\begin{figure*}[!hb]
\centering
     \includegraphics[width=\textwidth,height=0.4\textheight,keepaspectratio]{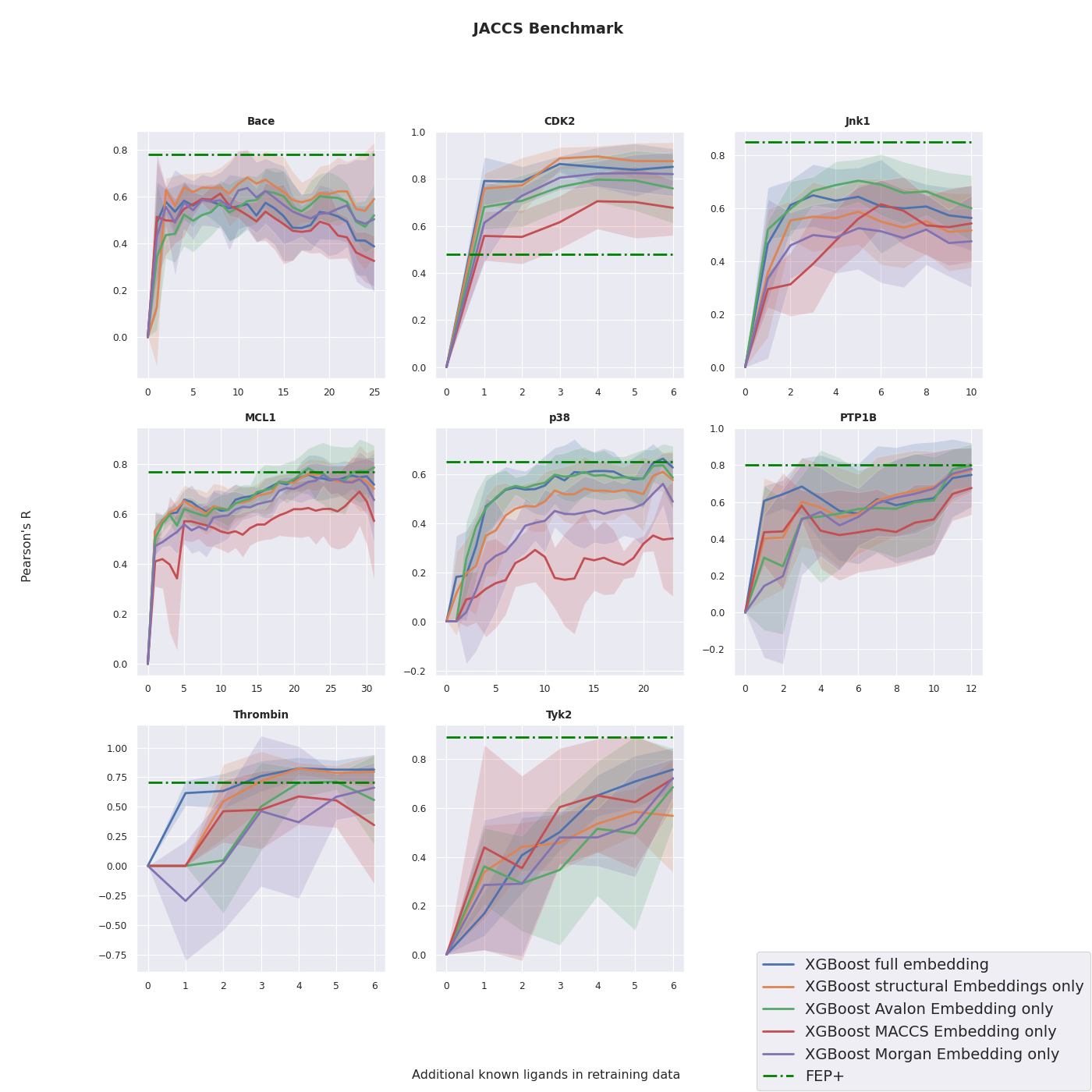}
      \caption{Learning curves comparing XGBoost models on the JACCS benchmark sets, trained using both combined and individual molecular RDKit embeddings. Plotted are Pearson's correlations against the pool of test compounds. The shaded area represents the standard deviation in the computed metric across the 25 model replicas. Included is also the performance of FEP+ on predicting absolute binding affinity as a reference.}
       \label{fig:schrodinger_bench_pears_rdkit_comp}
\end{figure*}

\begin{figure*}[!ht]
\centering
     \includegraphics[width=\textwidth,height=0.4\textheight,keepaspectratio]{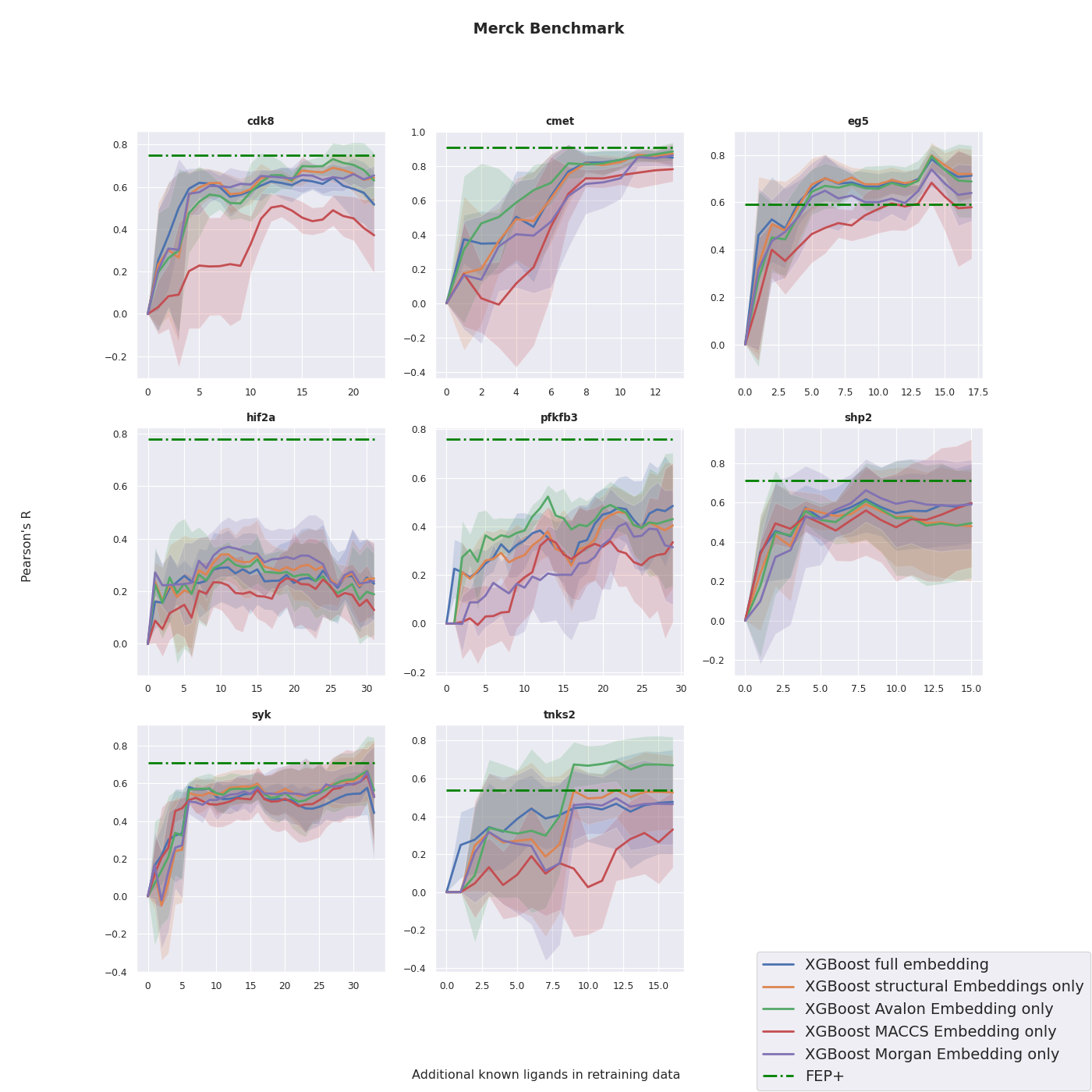}
      \caption{Learning curves for the Merck benchmark sets comparing XGBoost models trained on combined and individual molecular RDKit embeddings, illustrating the effect of different embedding strategies on model performance. Plotted are Pearson's correlations against the pool of test compounds. The shaded area represents the standard deviation in the computed metric across the 25 model replicas. Included is also the performance of FEP+ on predicting absolute binding affinity as a reference.}
       \label{fig:merck_bench_pears_rdkit_comp}
\end{figure*}

For integration into the XGBoost model, which requires fixed-dimension inputs, sum pooling operations were conducted on the per-token embeddings from the GPT2 model. Unlike GPT2, which uses attention mechanisms focused on preceding tokens, BERT incorporates a self-attention mechanism that covers all tokens, providing a comprehensive global embedding for the entire SMILES string. This is achieved through a special regression token that aggregates information from all tokens. To ensure a fair comparison of the head models mapping LLM embeddings to binding affinity values, we also trained an XGBoost model using the per-token BioT5+ embeddings, applying a similar sum pooling operation as used with the GPT2 embeddings.

From the results in figure \ref{fig:schrodinger_bench_pears_extra}, we observe that across various targets, the XGBoost model generally underperforms compared to the transformer model when both use BioT5+ embeddings, often exhibiting larger errors. This underperformance can be attributed to the transformer architecture's flexible, per-token attention mechanism, which effectively extracts information from the BioT5+ embedding. In contrast, the sum pooling operation employed with the XGBoost model averages out the per-token details, which complicates the extraction of necessary information and likely contributes to the observed lower performance. This also affects the performance of the GPT2 embeddings and can explain their lower performance on some targets.

The performance of the BERT embeddings, on the other hand, shows variability among different targets when compared to the BioT5+ embeddings. Since the BERT model does not undergo additional pooling operations, its fluctuating performance is likely not attributable to the XGBoost model head but rather to the smaller dataset used for its pretraining, unlike the more extensively pretrained BioT5+ model.

Beyond performance metrics, it is important to address several advantages that LLM-generated molecular embeddings have over traditional fingerprints, such as those found in RDKit. Due to their flexibility, LLMs can learn and embed a broad spectrum of information, making such embeddings highly versatile and adaptable for a variety of chemical and biomedical applications. These embeddings can also be fine-tuned to meet specific prediction task requirements. In contrast, RDKit embeddings statically encode specific information, with no capability for subsequent modification without altering the embedding strategy. LLMs are also able to embed the molecular information into a more compact form, which is highlighted by the lower embedding dimensions of the LLM-based embeddings. This is particularly usefull for the construction and storage of large molecular libraries in vector form from which molecules can be further selected based on their estimated binding affinities against a given target. Additionally, the ability to perform LLM embedding of molecular libraries on GPUs significantly accelerates library preparation compared to CPU-based RDKit preparation, offering substantial speed improvements. To enhance the speed of RDKit molecular embeddings, efforts like the CDPKit package \cite{cdpkit} have been initiated. However, CDPKit only includes an implementation of the Morgan fingerprint embedding. To further investigate the impact of each structural fingerprint on model performance, we conducted additional benchmarks on the JACCS and Merck test sets. We used XGBoost models trained individually on each of the structural embeddings from RDKit utilized in this study to assess their effectiveness in predicting binding affinity.

From figures \ref{fig:schrodinger_bench_pears_rdkit_comp} and \ref{fig:merck_bench_pears_rdkit_comp}, it is evident that using the full RDKit embedding—which combines chemical and three structural fingerprints—performs equally well as training without the chemical information. This suggests that the three structural fingerprints collectively provide sufficient meaningful information for the predictive task. Upon analyzing the performance of each individual embedding, it is apparent that the MACCS keys generally exhibit the lowest performance across most targets. In contrast, the Morgan fingerprint displays varying performance across targets when compared to the combined fingerprint vectors. Notably, the Avalon fingerprint vector consistently mirrors the performance of the combined fingerprint vectors and even surpasses them for some targets. These variations in performance across different targets underscore the potential advantages of more flexible embedding approaches, such as those offered by LLMs. When pretrained with a diverse and ample dataset, LLMs can generate embeddings that are more generalizable to various targets and downstream tasks.

For all plots presented in this section, additional figures covering other evaluation metrics, such as mean absolute error and Spearman's correlation, are available in SI.

\subsubsection{Large Congeneric Series Set}\label{sec:large_congeneric_set}

We also conducted a benchmark study using a larger congeneric series \cite{al1} of 10,000 molecules against the Tyk2 target. Each molecule's binding affinity data was computed using an FEP-like method. In their analysis, the authors explored the performance of various ML models, selection strategies, and the impact of the number of ligands selected per round. They found that employing either Gaussian Process or Gradient Boosted Trees models, combined with a greedy selection strategy where the compounds with the highest binding affinity are selected during each round, yielded the highest recall of the top 1\% of active molecules. Additionally, they noted that increasing the number of molecules selected during each round improved recall. However, practical implementations of active learning cycles must balance the number of selected molecules with the feasibility of screening these selections using computationally demanding methods such as ATM \cite{atm,atm2,atm3,atm4}, or through wet-lab experimental validation. This balance is necessary to provide a sufficient amount of new data to finetune the ML model during each cycle while keeping the costly experimental validation or use of computationally more demanding methods at a minimum.

To implement a feasible approach in a real-world active learning scenario, we opted for the same greedy selection strategy with a selection size of 60 molecules per round, aiming for a balance that could realistically be maintained and that showed good previous results \cite{al1}. Initially, we benchmarked our XGBoost model, which utilized the full RDKit embedding, against both a random baseline—where molecules were selected randomly without any ML model—and the best-performing Gaussian Process Regression model from \cite{al1}, under identical selection conditions. Five model replicas were trained during each round, and used to screen the remaining compounds in order to select molecules with the highest predicted binding affinities.

Typically, when initiating an active learning cycle, whether for a new target, binding pocket, or congeneric series, it is common to start with a random initial selection of molecule candidates \cite{al1}, where their binding affinity is determined using accurate but computationally demanding methods such as FEP+ or ATM or through experimental validation. To explore more effective initial selection strategies, we compared the conventional random selection to an ML-powered selection that leverages the capabilities of pretrained 3D models.

\begin{figure}[H]
\centering
     \includegraphics[width=1.0\columnwidth]{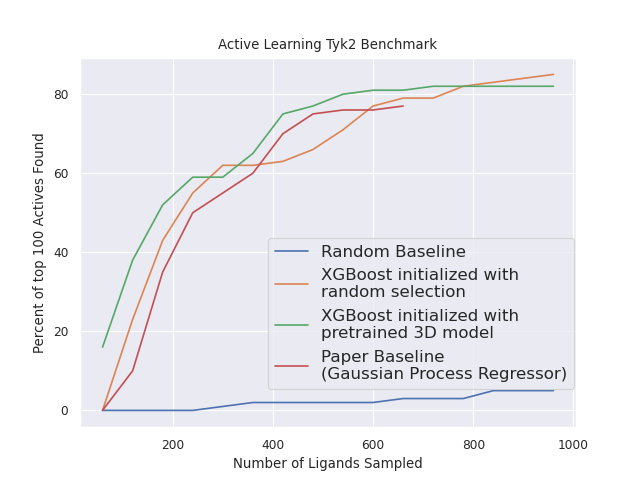}
     \captionsetup{width=1.0\columnwidth}
      \caption{Learning curve for the Tyk2 benchmark set: displayed is the percentage of top 1\% binders found after each iteration of the active learning cycle. It compares the performance of the XGBoost model using RDKit embeddings against two baselines: a non-ML random selection strategy and the best-performing model from \cite{al1}. Additionally, two initialization strategies for active learning are evaluated: a random selection of starting molecules and a targeted selection using the highest binding affinity predictions from a 3D graph neural network employing the TensorNet architecture. This network was pretrained on general binding affinity data, as detailed in section \ref{sec:gen_aff}.}
       \label{fig:large_tyk2_benchmark}
\end{figure}

From figure \ref{fig:large_tyk2_benchmark}, it is evident that our approach achieves a recall performance comparable to that of the best-performing model from the reference paper, under identical initialization and selection conditions and better than a completely random baseline. This highlights the effectiveness of ML-based screening during active learning.

Further observations reveal that by utilizing general binding affinity 3D models pretrained on heterogeneous binding affinity data (as described in section \ref{sec:gen_aff}), we can achieve a more effective initial selection of starting molecules. Thanks to the generalizability of these models, we are able to obtain a notable recall of top binding compounds even before the start of the active learning cycles. As the ML model's performance during active learning is influenced by the training data which depends on the selection of compounds made during the previous cycles, this improved initial selection also gives higher performance to the 2D model during subsequent cycles, showing to surpass the Gaussian model from the reference paper \cite{al1} in the recall of top binding compounds primarily during early cycles.

Furthermore, we conducted comparisons between XGBoost models, the 3D TensorNet model, and the BioT5+ LLM embeddings with a transformer head model. For the 3D model, compounds were docked against the Tyk2 target as described in section \ref{sec:test_datasets}. The transformer model, initiated with random weights, was finetuned during each iteration with the growing pool of training ligands, using early stopping when no improvement in training loss was observed for 50 steps. The TensorNet model, used for initializing the active learning cycles, was pretrained on a general binding affinity dataset as described in section \ref{sec:train_datasets} and subsequently finetuned for 300 epochs during each iteration. For each model, again five replicas were trained or finetuned in every cycle to ensure robustness and consistency in performance evaluations.

\begin{figure}[H]
\centering
     \includegraphics[width=1.0\columnwidth]{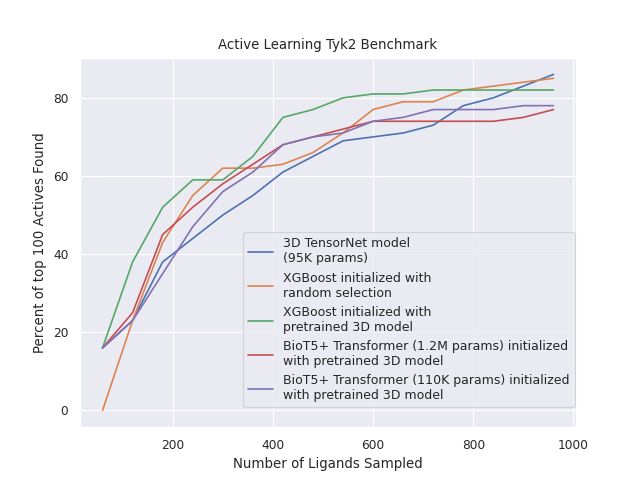}
     \captionsetup{width=1.0\columnwidth}
      \caption{Learning curve for the Tyk2 benchmark set: displayed is the percentage of top 1\% binders found after each iteration of the active learning cycle. The performance of XGBoost models, initialized with different molecule selection strategies, are compared to two neural network-based models: a 3D graph message passing network pretrained on general binding affinity data (as described in section \ref{sec:gen_aff}), and a transformer model utilizing molecular embeddings from the BioT5+ LLM. To ensure a fair comparison in terms of model complexity, two sizes of transformer models are tested against the 3D model.}
       \label{fig:large_tyk2_benchmark_models}
\end{figure}

From figure \ref{fig:large_tyk2_benchmark_models}, it is evident that the XGBoost model initialized with molecules from a pre-screening with the 3D model outperforms other models, identifying approximately 80\% of the top 1\% binders after sampling around 8\% of the data. In the early cycles, the neural network models show that pretrained LLM embeddings yield slightly better predictions than the 3D TensorNet model. This advantage is likely due to the LLM's more extensive pretraining dataset, which provides a significant benefit in low-data scenarios during the finetuning stages.

Considering the differences in model complexity, we included a comparison between the TensorNet model and a transformer model with fewer parameters. The results demonstrate that the smaller transformer model performs comparably to its larger counterpart and outperforms the 3D TensorNet model. Notably, the performance of the 3D model improves in the later stages of active learning relative to other models, underscoring the enhanced capabilities of neural network-based models with larger datasets.

To further explore the effect of certain hyperparameters on the performance of transformer-based models, results for the lighter version of the transformer model are presented in SI. These results examine different ratios of embedding size to the number of attention heads, emphasizing the critical role of this hyperparameter in optimizing transformer-based models.

\begin{figure}[H]
\centering
     \includegraphics[width=1.0\columnwidth]{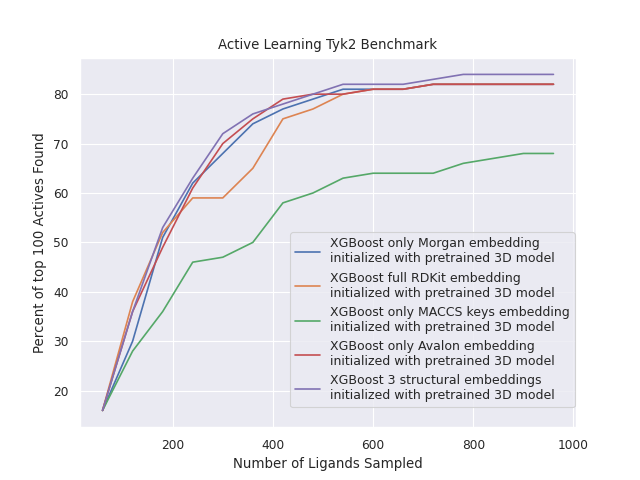}
     \captionsetup{width=1.0\columnwidth}
      \caption{Learning curve for the Tyk2 benchmark set: This chart plots the percentage of top 1\% binders found after each iteration of the active learning cycle. It compares the performance of XGBoost models trained on different RDKit embeddings: the full RDKit embedding, only the three structural RDKit fingerprints (Morgan, MACCS, and Avalon), and each of the three structural fingerprints individually. All models initiated the active learning with molecules selected by the pretrained 3D model.}
       \label{fig:large_tyk2_benchmark_rdkit_fps}
\end{figure}

Similarly to the JACCS and Merck benchmark sets, we also compared the performance of XGBoost models trained on each individual structural RDKit embedding. As observed previously in the JACCS and Merck benchmarks, figure \ref{fig:large_tyk2_benchmark_rdkit_fps} demonstrates that training XGBoost models on MACCS keys alone again results in the poorest performance. Furthermore, training on a combination of the three structural fingerprints or using Morgan or Avalon fingerprints individually tends to yield slightly better performance than including chemical information in the embedding, particularly in the mid-stages of the active learning cycle.

\subsection{Classifier models}

Given that the models in section \ref{sec:gen_aff} were trained exclusively on binders, they may struggle to identify non-binding molecules in a virtual screening scenario, often assigning binding affinity values to compounds that would not actually bind to a specific target. This limitation stems from the absence of non-binding molecules in the training dataset and the ability of docking programs to generate convincing docking poses for non-binders.

To address this issue without adversely affecting the regression accuracy on experimentally obtained binding affinity values, we transformed the GraphNet model architecture into a binary classifier. This adaptation involved updating the loss function to cross-entropy, enabling the model to output a probability of a ligand being a binder. However, in order to ensure the classifier does not merely learn structural or physicochemical biases from the training set, which could simplify the prediction task, we used the augmented training set from the decoy experiment described in section \ref{sec:gen_aff}. In this training set, binders were labeled "1" and decoys "0". Considering that a tree-based model without explicit protein context performed poorly on such data and to ensure the classifier's generalizability, especially in scenarios where the same ligand might bind to certain targets but not others, we opted not to use a tree-based model for this task.

We evaluated the classifier's performance through 5-fold cross-validation on the training set. Each fold was created by assigning entire protein clusters to different folds, testing the model’s generalizability on unseen and structurally different targets than those in the training data. Protein clusters were obtained using BLAST with a similarity cutoff of 0.7.

\begin{figure}[H]
\centering
     \includegraphics[width=1.0\columnwidth]{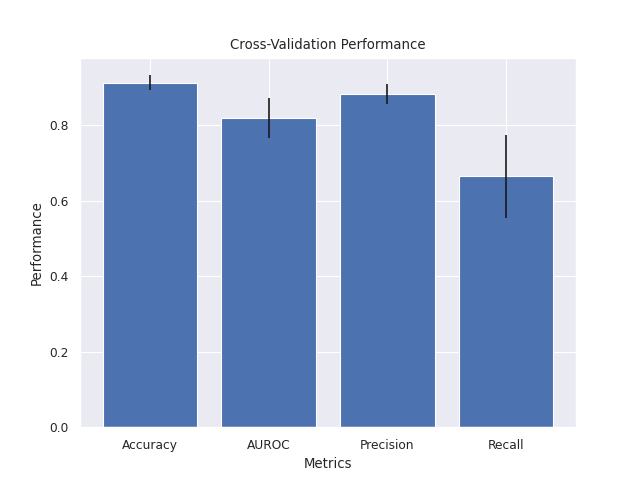}
     \captionsetup{width=1.0\columnwidth}
      \caption{Classification performance from 5-fold cross-validation, averaged over the 5 folds. Black lines indicate the standard deviation across the folds.}
       \label{fig:class_performance}
\end{figure}

We calculated several classification metrics to assess the model's accuracy. From the plots in figure \ref{fig:class_performance}, it is evident that the model demonstrates good accuracy in classifying compounds. However, it shows a slight bias toward the decoy class, indicated by a higher false negative rate. Despite this, the model maintains a high true positive rate. Given that the primary goal of the model is to ensure that predicted binders are indeed true binders, we consider this performance to be acceptable.

\section{Conclusion}

This study assessed graph-based 3D and tree-based 2D models for binding affinity prediction across various predictive modalities, benchmarking them against well-established benchmark datasets. While graph models did not consistently surpass simpler ligand-only models, they proved more effective if the task was not just to rank good binders but also differentiate bad binders from decoys especially when both are structurally and physicochemically similar. Additionally, tree-based models demonstrated particular strength during active learning cycles, exhibiting resistance to overfitting despite limited data. In contrast, neural network models displayed matching performance towards the late stages of active learning with more available data points. Supervised pretraining of 3D graph neural network models enhanced their performance during early and active learning cycles, highlighting their reliance on larger training datasets.

Given these results, we also tested if larger, unsupervised pretraining could further improve the performance of neural network-based models. While unsupervised pretraining of 3D graph-based neural network models showed interesting results \cite{denoising}, in this work, we tested unsupervised pretraining of large language models (LLMs). Molecular embeddings obtained from these pretrained LLMs demonstrated comparable performance to classical RDKit embeddings and improvement over 3D graph-based neural network models pretrained in a supervised fashion across several active learning scenarios, underscoring the benefits of large-scale unsupervised pretraining. Given the advantages of LLM embeddings—including richer information content, a more compact nature, and faster generation—proper pretraining tailored to the downstream binding affinity prediction task could allow LLM embeddings to be competitive with traditional embedding methods.

Finally, we explored the optimal applications of both 3D and 2D models. 3D models, capable of learning from diverse data, prove beneficial for general binding affinity prediction tasks, such as initial library screenings in early hit discovery or initiating active learning cycles, given unsupervised and/or supervised pretraining on large and diverse datasets, augmented to suppress data-specific biases. Conversely, 2D models are more effective in low-data regimes, typically seen in the early stages of active learning cycles. By leveraging the strengths of both 3D and 2D models, we showed how to achieve state-of-the-art results in simulated active learning scenarios.

\section{Supporting Information}

Supporting Information: "congeneric series test sets descriptions", "optimized models' hyperparameter values" and "additional results", additional information for the sections; Table S1 information on congeneric series test sets compositions; Tables S2-S6 optimal hyperparameter configurations for tested models; Figures S1-S2 additional performance metrics for general binding affinity models and benchmarks; Figures S3-S4 additional performance metrics for general binding affinity models and benchmarks using decoys; Figures S5-S6 additional performance metrics for general binding affinity models and benchmarks with crystal waters; Figures S7-S14 additional learning curves on multi-scaffold ligand series for other targets; Figures S15-S16 additional performance metrics JACCS congeneric series LLM transformer and XGBoost models; Figures S17-S18 additional performance metrics Merck congeneric series LLM transformer and XGBoost models; Figures S19-S20 additional performance metrics JACCS congeneric series LLM embeddings comparisons; Figures S21-S22 additional performance metrics JACCS congeneric series RDKit embeddings comparisons; Figures S23-S24 additional performance metrics Merck congeneric series RDKit embeddings comparisons; Figure S25 attention heads transformer model hyperparameter test on large congeneric series set (PDF)

\section{Acknowledgements}

This work was supported by the Industrial Doctorates Plan of the Secretariat of Universities and Research of the Department of Economy and Knowledge of the Generalitat of Catalonia.

\section{Data and Software Availability}

All datasets used in this work come from public sources. The PDBBind data, described in \cite{pdbbind}, can be downloaded from \hyperlink{http://www.pdbbind.org.cn/}{pdbbind.org.cn}. BindingDB data is described in \cite{bindingdb} and available through \hyperlink{https://www.bindingdb.org}{bindingdb.org} with the validation sets available through the \hyperlink{https://www.bindingdb.org/rwd/validation_sets/index.jsp}{Special Datasets} subsection. Additionally the prepared training data is also available from our KDeepTrainer application available via \hyperlink{https://open.playmolecule.org}{PlayMolecule.org}. The JACCS benchmark sets are available through \hyperlink{https://github.com/schrodinger/public_binding_free_energy_benchmark?tab=readme-ov-file}{https://github.com/schrodinger} and are described further in \cite{fep_schrodinger_set}. The Merck benchmark sets are available through \hyperlink{https://github.com/MCompChem/fep-benchmark}{https://github.com/MCompChem/fep-benchmark} and are described in \cite{merck}. The OpenForceField benchmark sets are available through \hyperlink{https://github.com/openforcefield/openff-benchmark}{https://github.com/openforcefield/openff-benchmark} and are described in \cite{openff}. Lastly, the large Tyk2 benchmark set is available through \hyperlink{https://github.com/google-research/google-research/tree/master/al_for_fep}{https://github.com/google-research/alforfep} and is described in \cite{al1}.
The 3D models can be trained or fine-tuned via the KDeepTrainer application and the pretrained models used in this work are all available through the application interface. Screening of datasets with pretrained 3D models can be done via the KDeep application. The 2D models are available for either training or screening of libraries through the KDeep2D application. In the latter application, users can specify whether to fit a model to provided training data or perform screening of provided molecule libraries with a trained model. Selection can be made from the application interface to use RDKit or LLM-based molecular embeddings. All the mentioned applications are publicly available via \hyperlink{https://open.playmolecule.org}{PlayMolecule.org}.

\printbibliography
\end{multicols}
\newpage
\tableofcontents

\end{document}

\maketitle

\tableofcontents
\addcontentsline{toc}{section}{List of Figures}
\listoffigures
\addcontentsline{toc}{section}{List of Tables}
\listoftables

\section{FEP Benchmark Datasets Overview}

% Please add the following required packages to your document preamble:
% \usepackage{graphicx}
\begin{table}[H]
\centering
\resizebox{0.7\columnwidth}{!}{%
\tiny
\begin{tabular}{ll}
Target       & Number of Ligands    \\
\multicolumn{2}{c}{JACCS Set} \\
Bace         & 36                   \\
CDK2         & 16                   \\
Jnk1         & 21                   \\
MCL1         & 42                   \\
p38          & 34                   \\
PTP1B        & 23                   \\
Thrombin     & 11                   \\
Tyk2         & 16                   \\
\multicolumn{2}{c}{Merck Set}       \\
cdk8         & 33                   \\
cmet         & 24                   \\
eg5          & 28                   \\
hif2a        & 42                   \\
pfkfb3       & 40                   \\
shp2         & 26                   \\
syk          & 44                   \\
tnks2        & 27                   \\
\multicolumn{2}{c}{OpenForceField Set}       \\
CDK2         & 10                   \\
CDK8         & 31                   \\
CMET         & 5                    \\
eg5          & 27                   \\
hif2a        & 37                   \\
MCL1         & 25                   \\
p38          & 29                   \\
PDE2         & 21                   \\
PFKFB3       & 32                   \\
PTP1B        & 22                   \\
SHP2         & 24                   \\
SYK          & 44                   \\
Thrombin     & 23                   \\
TNKS2        & 27                   \\
TYK2         & 13                   \\
\end{tabular}%
}
\caption{Overview of Benchmark Test Sets: This table lists the composition of each test set, detailing the targets and the number of ligands in each ligand series included in the study.}
\label{tab:fep_test_sets}
\end{table}

\section{Models' Hyperparameters}

\begin{table}[H]
\begin{tabular}{ll}
\textbf{Hyperparameter}                                                    & \textbf{Value}    \\ \hline
Embedding dimension                                                        & 64                \\
Number of distance functions                                               & 64                \\
Distance function                                                          & Gaussian          \\
\begin{tabular}[c]{@{}l@{}}Number of message passing\\ layers\end{tabular} & 6                 \\
Lower atom cutoff                                                          & 0                 \\
Upper atom cutoff                                                          & 6                 \\
Initial learning rate                                                      & 0.00075           \\
Learning rate scheduler                                                    & ReduceLROnPlateau \\
Scheduler factor                                                           & 0.9               \\
Scheduler patience                                                         & 26000 steps       \\
Activation function                                                        & SiLU             
\end{tabular}
\caption{Chosen Hyperparameters for the GraphNet Model: This table displays the hyperparameters selected for the GraphNet model. These parameters were consistently applied across the classification model as well, with the sole difference being the adoption of BinaryCrossEntropyLoss for the classification tasks.}
\label{tab:hyperparams_graphnet}
\end{table}

\begin{table}[H]
\begin{tabular}{ll}
\textbf{Hyperparameter}                                                    & \textbf{Value}                   \\ \hline
Embedding dimension                                                        & 64                               \\
Number of distance functions                                               & 64                               \\
Distance function                                                          & Gaussian                         \\
\begin{tabular}[c]{@{}l@{}}Number of message passing\\ layers\end{tabular} & 5                                \\
Lower atom cutoff                                                          & 0                                \\
Upper atom cutoff                                                          & 6                                \\
Initial learning rate                                                      & 0.00003                          \\
Learning rate scheduler                                                    & Slanted Triangular Learning Rate \\
Learning rate warmup steps                                                 & 80000                            \\
Scheduler factor                                                           & 0.9                              \\
Scheduler patience                                                         & 52000 steps                      \\
Activation function                                                        & SiLU                             \\
Number of attention heads                                                  & 8                                \\
Neighbor embedding                                                         & True                            
\end{tabular}
\caption{Chosen Hyperparameters for the Equivariant Transformer Model: This table lists the specific hyperparameters selected for the Equivariant Transformer model, detailing the settings used to optimize its performance.}
\label{tab:hyperparams_et}
\end{table}

\begin{table}[H]
\begin{tabular}{ll}
\textbf{Hyperparameter}                                                    & \textbf{Value}    \\ \hline
Embedding dimension                                                        & 64                \\
Number of distance functions                                               & 64                \\
Distance function                                                          & Gaussian          \\
\begin{tabular}[c]{@{}l@{}}Number of message passing\\ layers\end{tabular} & 0                 \\
Lower atom cutoff                                                          & 0                 \\
Upper atom cutoff                                                          & 9                 \\
Initial learning rate                                                      & 0.001             \\
Learning rate scheduler                                                    & ReduceLROnPlateau \\
Scheduler factor                                                           & 0.75              \\
Scheduler patience                                                         & 52000 steps       \\
Activation function                                                        & SiLU              \\
Neighbor embedding                                                         & True              \\
Number of linear scalars                                                   & 2                 \\
Number of linear tensors                                                   & 2                 \\
Loss smoothing                                                             & True             
\end{tabular}
\caption{Chosen Hyperparameters for the TensorNet Model: This table details the specific hyperparameters selected for the TensorNet model, outlining the settings that are critical for its optimal performance.}
\label{tab:hyperparams_tensornet}
\end{table}

\begin{table}[H]
\begin{tabular}{ll}
\textbf{Hyperparameter} & \textbf{Value} \\ \hline
Fingerprint radius      & 10             \\
Fingerprint length      & 4096           \\
Number of trees         & 20000          \\
Learning rate           & 0.001          \\
Maximum tree depth      & 7              \\
Subsample               & 0.3            \\
Descriptor types & \begin{tabular}[c]{@{}l@{}}Chemical, Morgan, \\ MACCS, Avalon\end{tabular}
\end{tabular}
\caption{Chosen Hyperparameters for the XGBoost Model: This table lists the specific hyperparameters selected for the XGBoost model, including details unique to different descriptors. The fingerprint radius applies only to the Morgan and MACCS descriptors, while the fingerprint length is relevant only for the Morgan and Avalon descriptors. The total embedding length for the full RDKit embedding is 8567.}
\label{tab:hyperparams_xgboost}
\end{table}

\begin{table}[H]
\begin{tabular}{ll}
\textbf{Hyperparameter}   & \textbf{Value}    \\ \hline
Token embedding size      & 256               \\
Number of layers          & 2                 \\
Number of attention heads & 16                \\
Initial learning rate     & 0.0001            \\
Learning rate scheduler   & ReduceLROnPlateau \\
Scheduler factor          & 0.9               \\
Scheduler patience        & 1500              \\
Early stopping delta      & 0.002             \\
Early stopping patience   & 50 steps         
\end{tabular}
\caption{Chosen Hyperparameters for the LLM Transformer Model: This table outlines the specific hyperparameters selected for the LLM Transformer model. Notably, the embedding lengths vary among the LLMs used; the BioT5+ LLM embedding is 786, while the GPT2 embedding is 256 and the BERT embedding is 128.}
\label{tab:hyperparams_transformer}
\end{table}

\section{Additional plots for Experiments and Results}
\subsection{General Binding Affinity Models}

\begin{figure}[H]
\centering
     \includegraphics[width=1.0\textwidth]{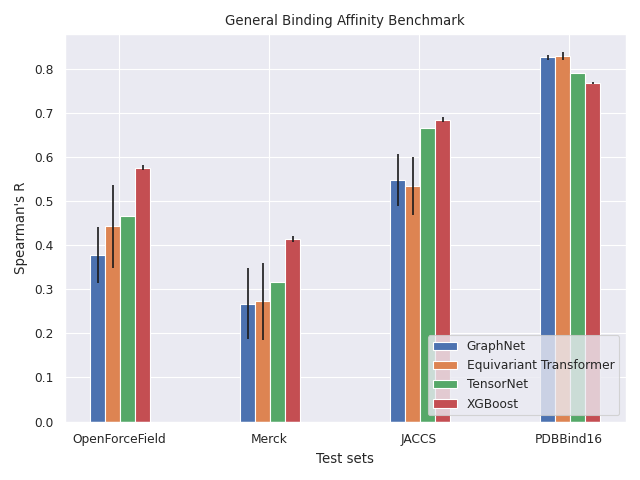}
      \caption{Benchmark of General Affinity Models, Spearman's R: This figure compares various model architectures trained on collected binding affinity data across different test sets. Eight replicas were trained for each model, and error bars illustrate the standard deviation of the predicted metric among these replicas. Due to extended training times, only one replica was trained for the TensorNet architecture.}
       \label{fig:bench_general_spearman}
\end{figure}

\begin{figure}[H]
\centering
     \includegraphics[width=1.0\textwidth]{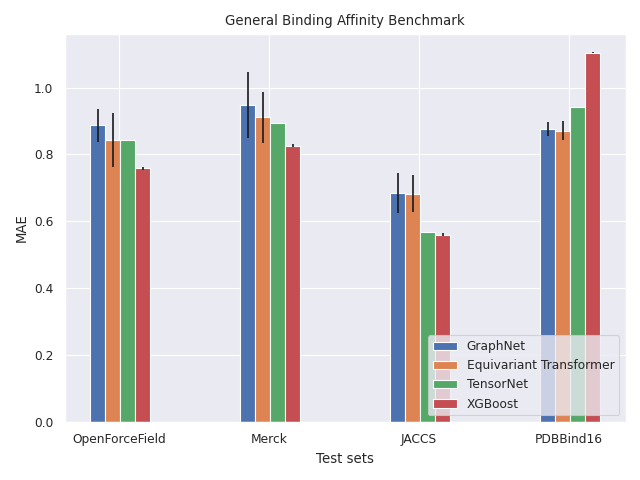}
      \caption{Benchmark of General Affinity Models, MAE: This figure displays the performance of various model architectures trained on collected binding affinity data and tested across different prepared test sets. Eight replicas were trained for each model, with error bars indicating the standard deviation of the predicted metric across these replicas. Notably, due to longer training durations, only one replica was trained for the TensorNet architecture.}
       \label{fig:bench_general_mae}
\end{figure}

\begin{figure}[H]
\centering
     \includegraphics[width=1.0\textwidth]{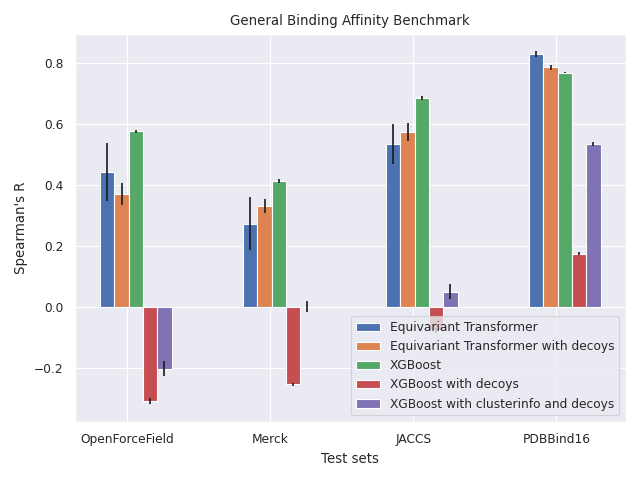}
      \caption{Benchmark of General Affinity Models with Decoys, Spearman's R: This figure compares the performance of the equivariant transformer model and the ligand-based XGBoost model, both tested with the addition of decoy molecules. For models incorporating decoys, three replicas were trained, whereas eight replicas were used for models without decoys. Error bars indicate the standard deviation of the predicted metric across these replicas.}
       \label{fig:bench_decoy_spearman}
\end{figure}

\begin{figure}[H]
\centering
     \includegraphics[width=1.0\textwidth]{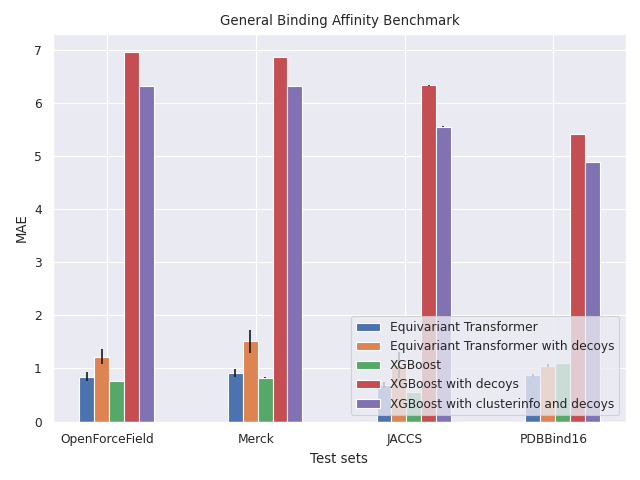}
      \caption{Benchmark of General Affinity Models with Decoys, MAE: This figure shows results from testing the equivariant transformer model and ligand-based XGBoost model with the addition of decoy molecules. Three replicas of each model were trained with decoys, while eight replicas were trained without decoys. Error bars illustrate the standard deviation of the predicted metric across these replicas.}
       \label{fig:bench_decoy_mae}
\end{figure}

\subsection{Influence of Waters on Performance of General Binding Affinity Models}

\begin{figure}[H]
\centering
     \includegraphics[width=1.0\textwidth]{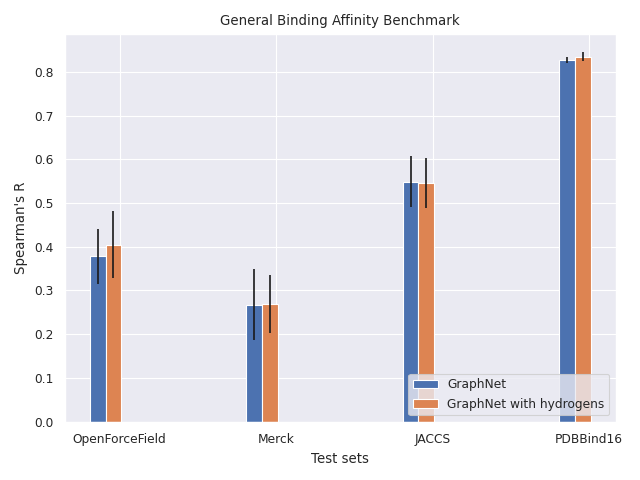}
      \caption{Comparison of the 3D GraphNet model performance when trained with and without crystal water and hydrogen atoms. The positions of both elements were sourced from the provided crystal structures and the prepared test sets. 8 model replicas were trained for each model and error bars represent standard deviation on the computed Spearman's correlation metric across model replicas.}
       \label{fig:bench_waters_spearman}
\end{figure}

\begin{figure}[H]
\centering
     \includegraphics[width=1.0\textwidth]{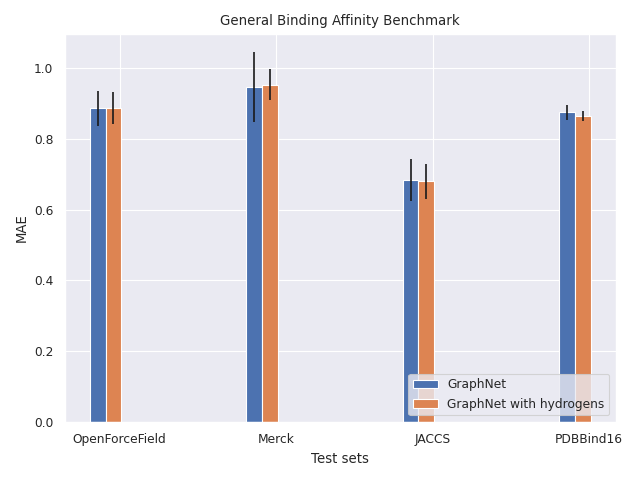}
      \caption{Comparison of the 3D GraphNet model performance when trained with and without crystal water and hydrogen atoms. The positions of both elements were sourced from the provided crystal structures and the prepared test sets. 8 model replicas were trained for each model and error bars represent standard deviation on the computed mean absolute error correlation metric across model replicas.}
       \label{fig:bench_waters_mae}
\end{figure}

\subsection{Additional Learning Curves for other Targets of Multi-Scaffold Sets}

\begin{figure}[H]
\centering
     \includegraphics[width=1.0\textwidth]{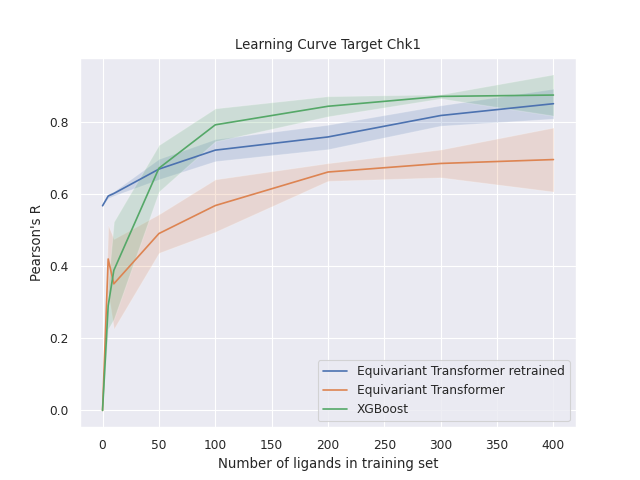}
      \caption{Learning Curve for Chk1 Target: This graph displays the progression of models performance over successive training iterations on the Chk1 target dataset, illustrating how predictive accuracy evolves during the active learning cycles. Shaded area represents standard deviation on the computed metric across model replicas.}
       \label{fig:lc_chk1}
\end{figure}

\begin{figure}[H]
\centering
     \includegraphics[width=1.0\textwidth]{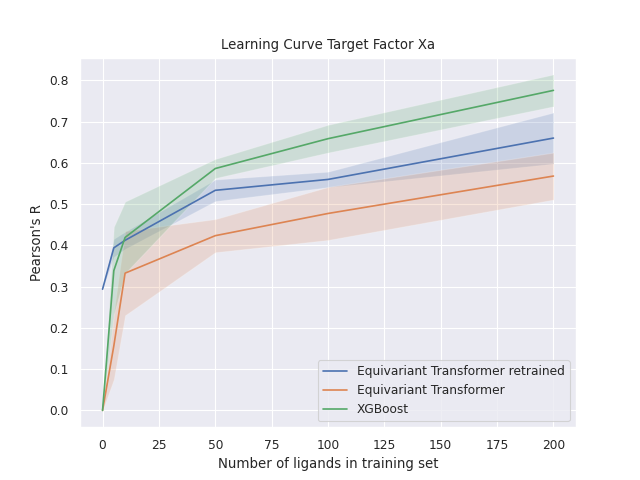}
      \caption{Learning Curve for Factor Xa Target: This graph displays the progression of models performance over successive training iterations on the Factor Xa target dataset, illustrating how predictive accuracy evolves during the active learning cycles. Shaded area represents standard deviation on the computed metric across model replicas.}
       \label{fig:lc_f10a}
\end{figure}

\begin{figure}[H]
\centering
     \includegraphics[width=1.0\textwidth]{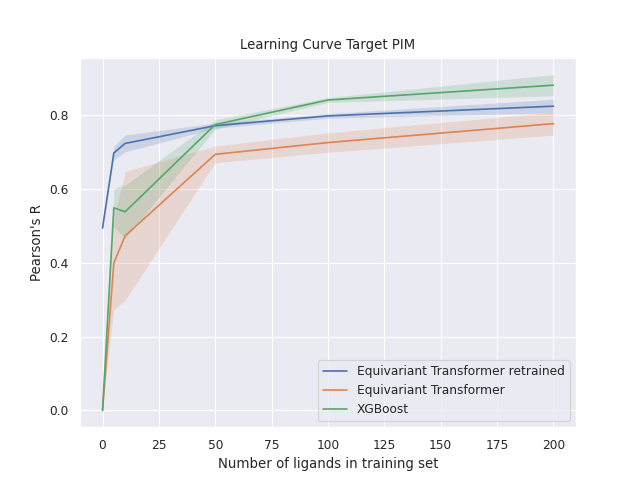}
      \caption{Learning Curve for PIM-I Target: This graph displays the progression of models performance over successive training iterations on the PIM-I target dataset, illustrating how predictive accuracy evolves during the active learning cycles. Shaded area represents standard deviation on the computed metric across model replicas.}
       \label{fig:lc_pim1}
\end{figure}

\begin{figure}[H]
\centering
     \includegraphics[width=1.0\textwidth]{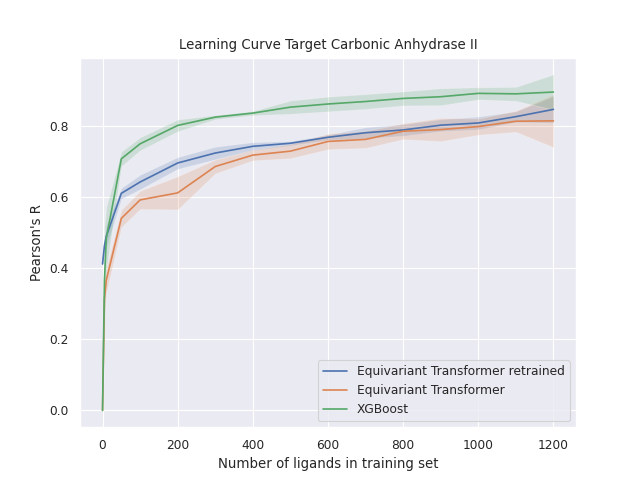}
      \caption{Learning Curve for Carbonic Anhydrase II Target: This graph displays the progression of models performance over successive training iterations on the Carbonic Anhydrase II target dataset, illustrating how predictive accuracy evolves during the active learning cycles. Shaded area represents standard deviation on the computed metric across model replicas.}
       \label{fig:lc_caII}
\end{figure}

\begin{figure}[H]
\centering
     \includegraphics[width=1.0\textwidth]{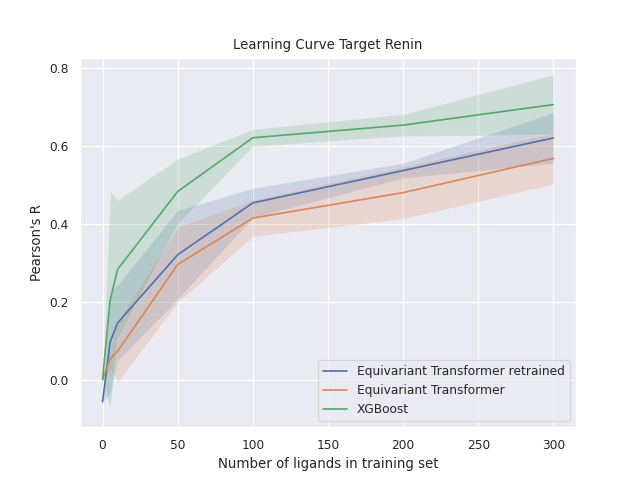}
      \caption{Learning Curve for Renin Target: This graph displays the progression of models performance over successive training iterations on the Renin target dataset, illustrating how predictive accuracy evolves during the active learning cycles. Shaded area represents standard deviation on the computed metric across model replicas.}
       \label{fig:lc_renin}
\end{figure}

\subsection{Additional metrics for JACCS Benchmark Sets}

\begin{figure}[H]
\centering
     \includegraphics[width=1.0\textwidth]{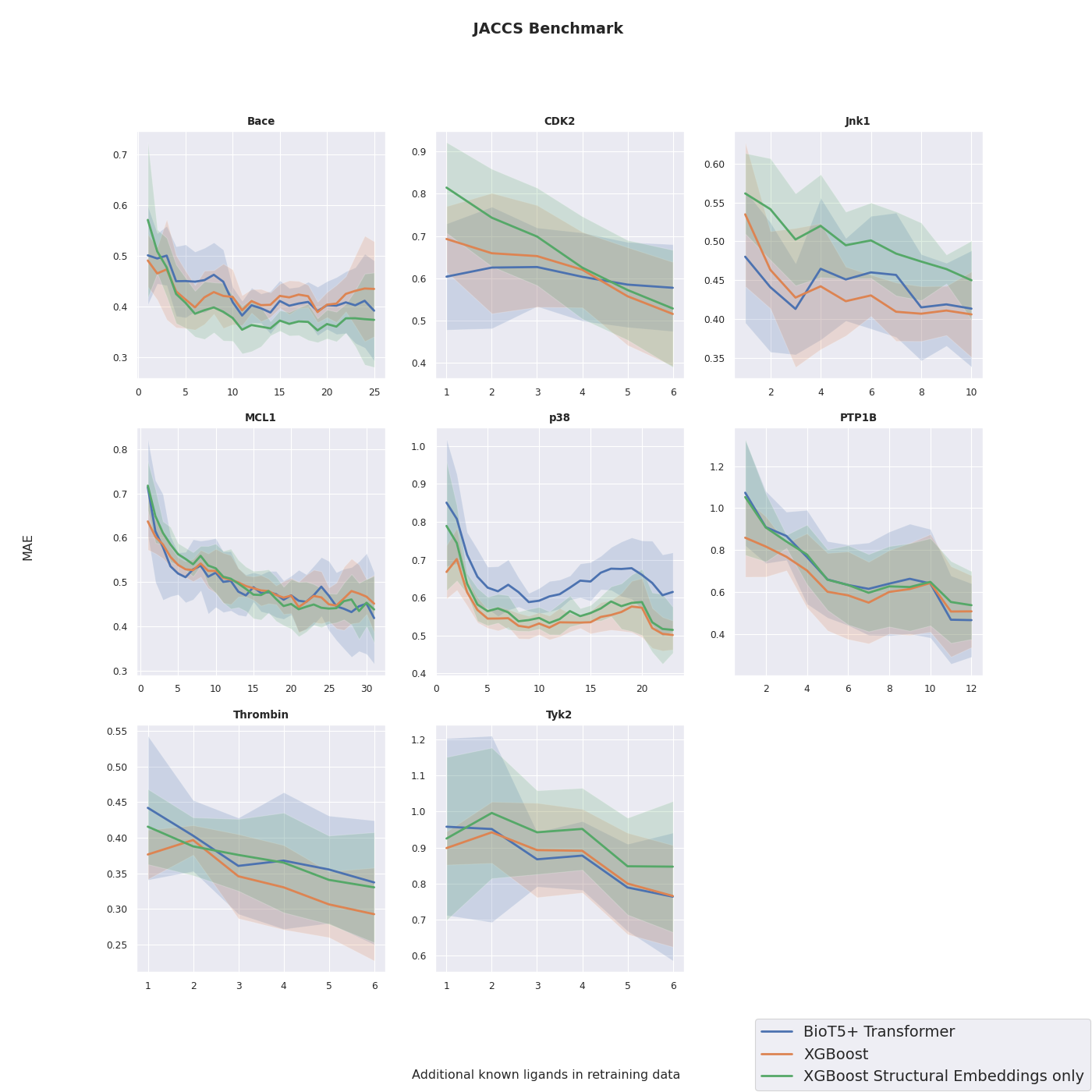}
      \caption{Learning Curves for JACCS Targets: This chart shows learning curves for several target test sets within the JACCS FEP benchmark sets. It compares the performance of an XGBoost model and a transformer model utilizing a pretrained LLM embedding that includes only structural information about the ligand. For additional comparison, results from an XGBoost model using only structural RDKit embeddings are also plotted. Plotted is the mean absolute error against the pool of test molecules. The shaded area represents the standard deviation in the computed metric across the 25 model replicas.}
       \label{fig:schrodinger_bench_mae}
\end{figure}

\begin{figure}[H]
\centering
     \includegraphics[width=1.0\textwidth]{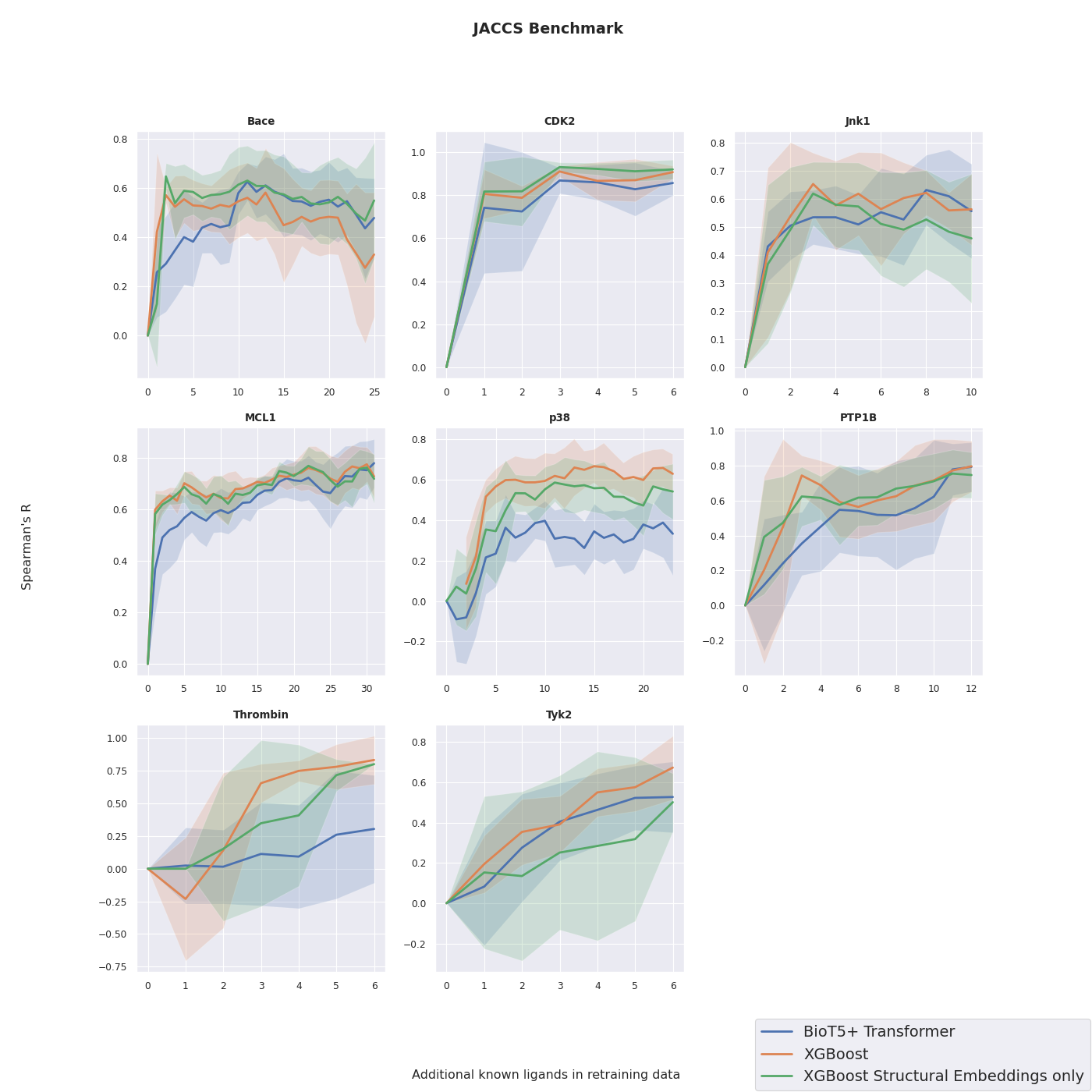}
      \caption{Learning Curves for JACCS Targets: This graph depicts learning curves for multiple target test sets within the JACCS FEP benchmark sets. The performance of an XGBoost model is compared with that of a transformer model using a pretrained LLM embedding, which focuses solely on the structural aspects of the ligand. Additionally, results from an XGBoost model employing only structural RDKit embeddings are included for further comparison. Plotted is the Spearman's correlation against the pool of test molecules. The shaded area represents the standard deviation in the computed metric across the 25 model replicas.}
       \label{fig:schrodinger_bench_spear}
\end{figure}

\subsection{Additional metrics for Merck Benchmark Sets}

\begin{figure}[H]
\centering
     \includegraphics[width=1.0\textwidth]{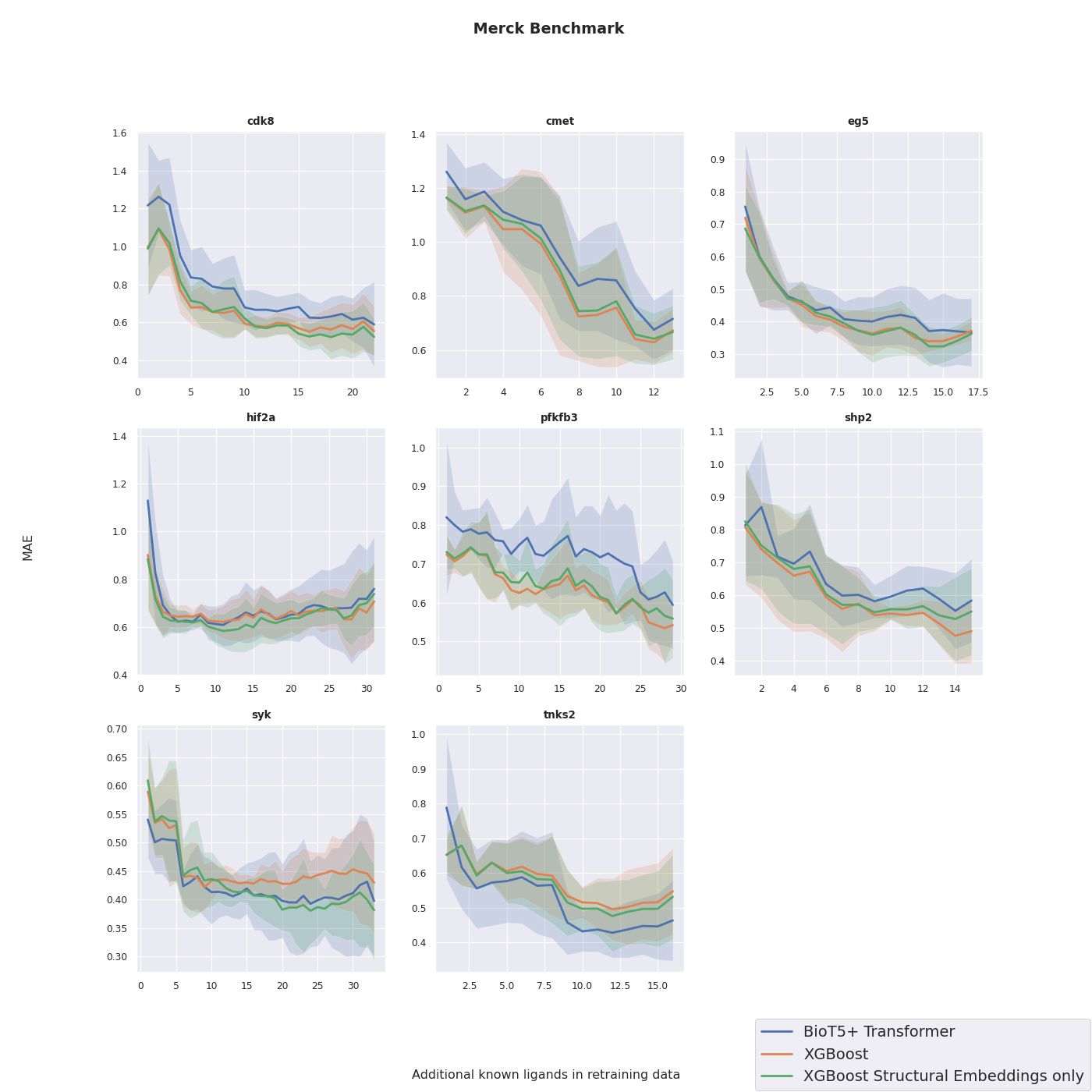}
      \caption{Learning Curves for Merck Targets: This graph illustrates learning curves for several target test sets within the Merck FEP benchmark sets. It compares the performance of an XGBoost model to that of a transformer model equipped with a pretrained LLM embedding, which focuses exclusively on structural information of the ligand. For additional comparison, performance data for an XGBoost model utilizing only structural RDKit embeddings is also displayed. Plotted is the mean absolute error against the pool of test molecules. The shaded area represents the standard deviation in the computed metric across the 25 model replicas.}
       \label{fig:merck_bench_mae}
\end{figure}

\begin{figure}[H]
\centering
     \includegraphics[width=1.0\textwidth]{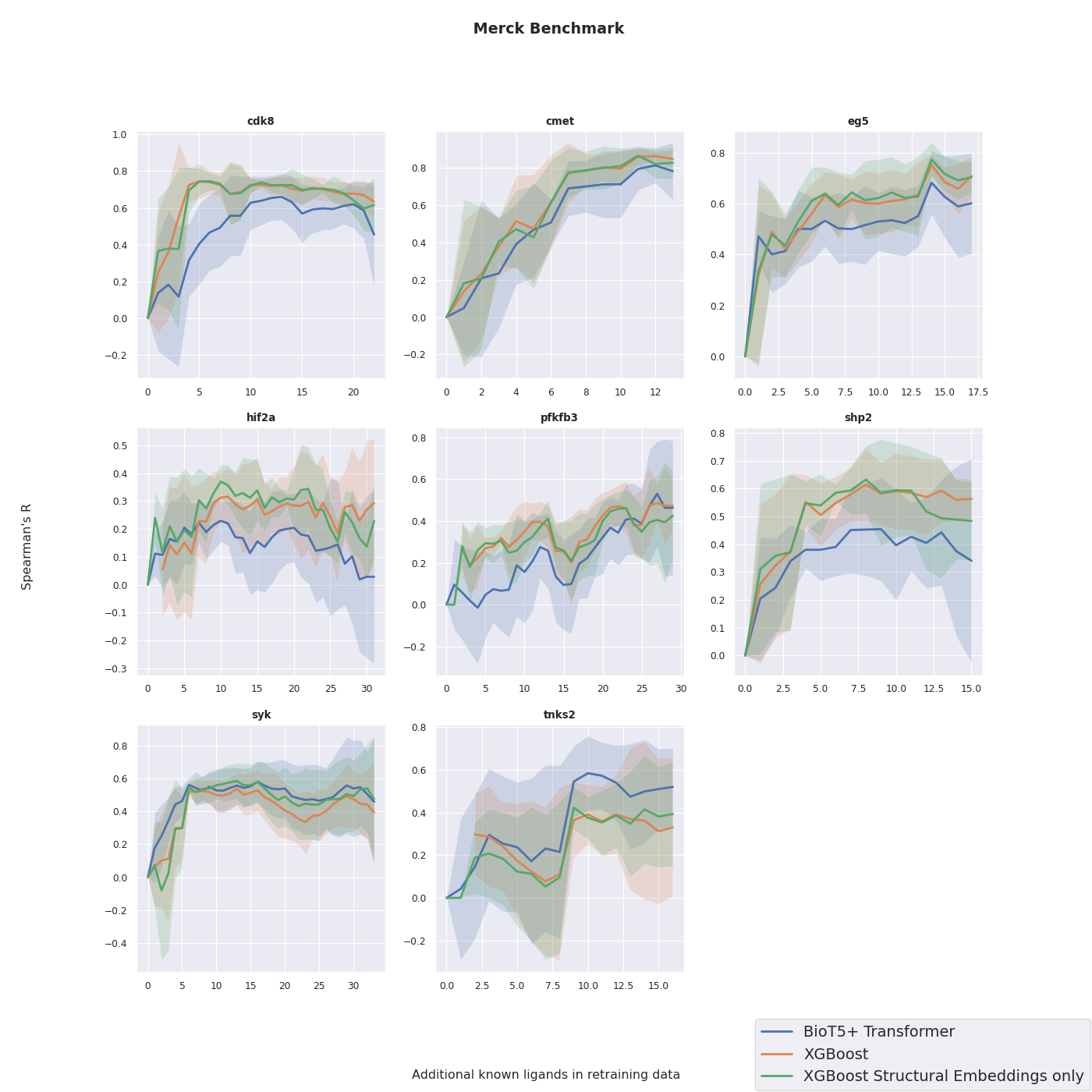}
      \caption{Learning Curves for Merck Targets: This figure displays learning curves for several target test sets within the Merck FEP benchmark sets. It features a comparison between the XGBoost model and a transformer model equipped with a pretrained LLM embedding that includes only structural information of the ligand. Additionally, results from an XGBoost model using solely structural RDKit embeddings are presented for comparison. Plotted is the Spearman's correlation against the pool of test molecules. The shaded area represents the standard deviation in the computed metric across the 25 model replicas.}
       \label{fig:merck_bench_spear}
\end{figure}

\subsection{Additional Plots Comparisons LLMs JACCS Benchmark}

\begin{figure}[H]
\centering
     \includegraphics[width=1.0\textwidth]{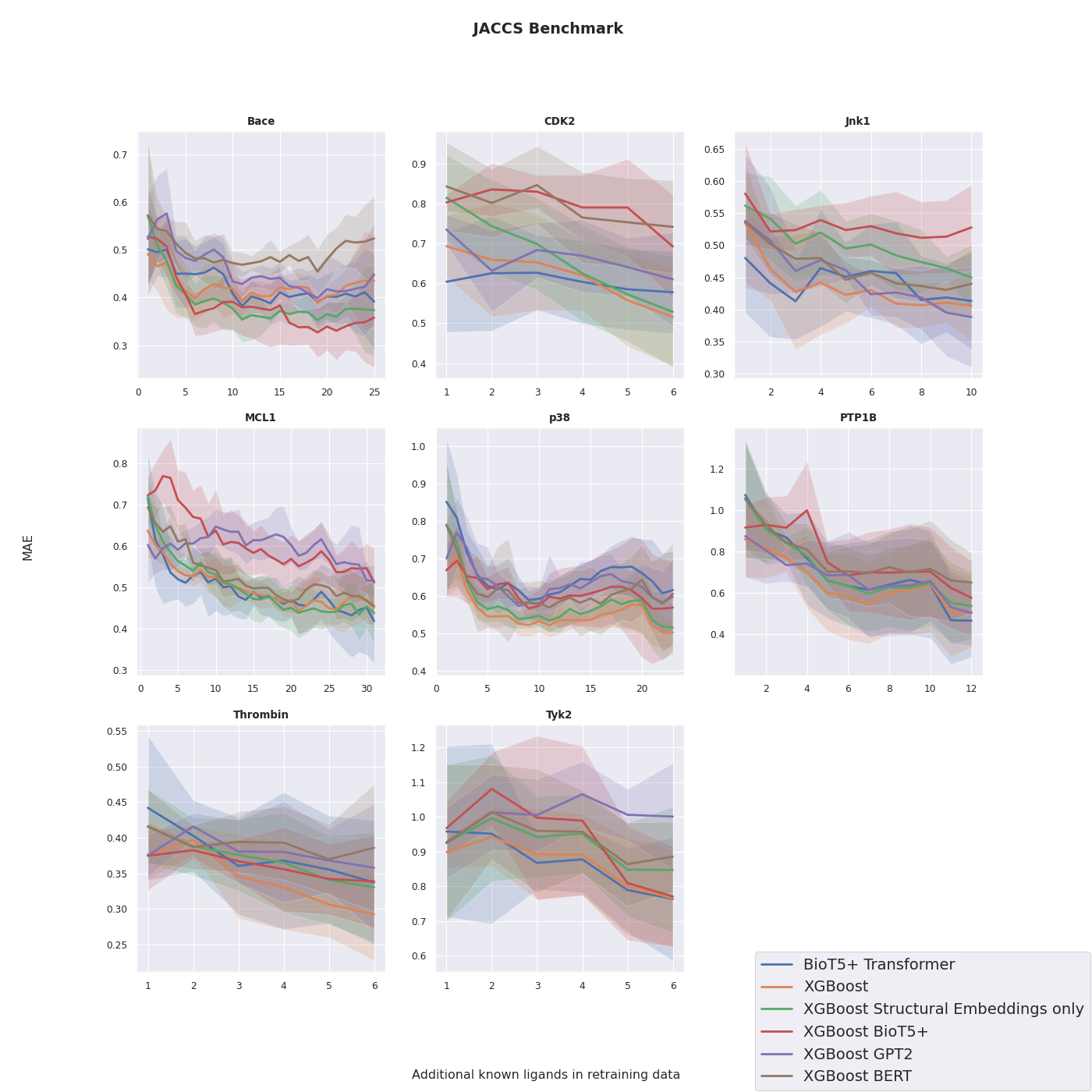}
      \caption{Additional Learning Curves on JACCS Benchmark Sets: This graph compares XGBoost models using RDKit embeddings against those utilizing LLM embeddings from various architectures. Given that LLM embeddings focus solely on structural information, the figure also includes performance data for an XGBoost model that employs only structural RDKit embeddings for a direct comparison. Plotted is the mean absolute error against the pool of test molecules. The shaded area represents the standard deviation in the computed metric across the 25 model replicas.}
       \label{fig:schrodinger_bench_mae_extra}
\end{figure}

\begin{figure}[H]
\centering
     \includegraphics[width=1.0\textwidth]{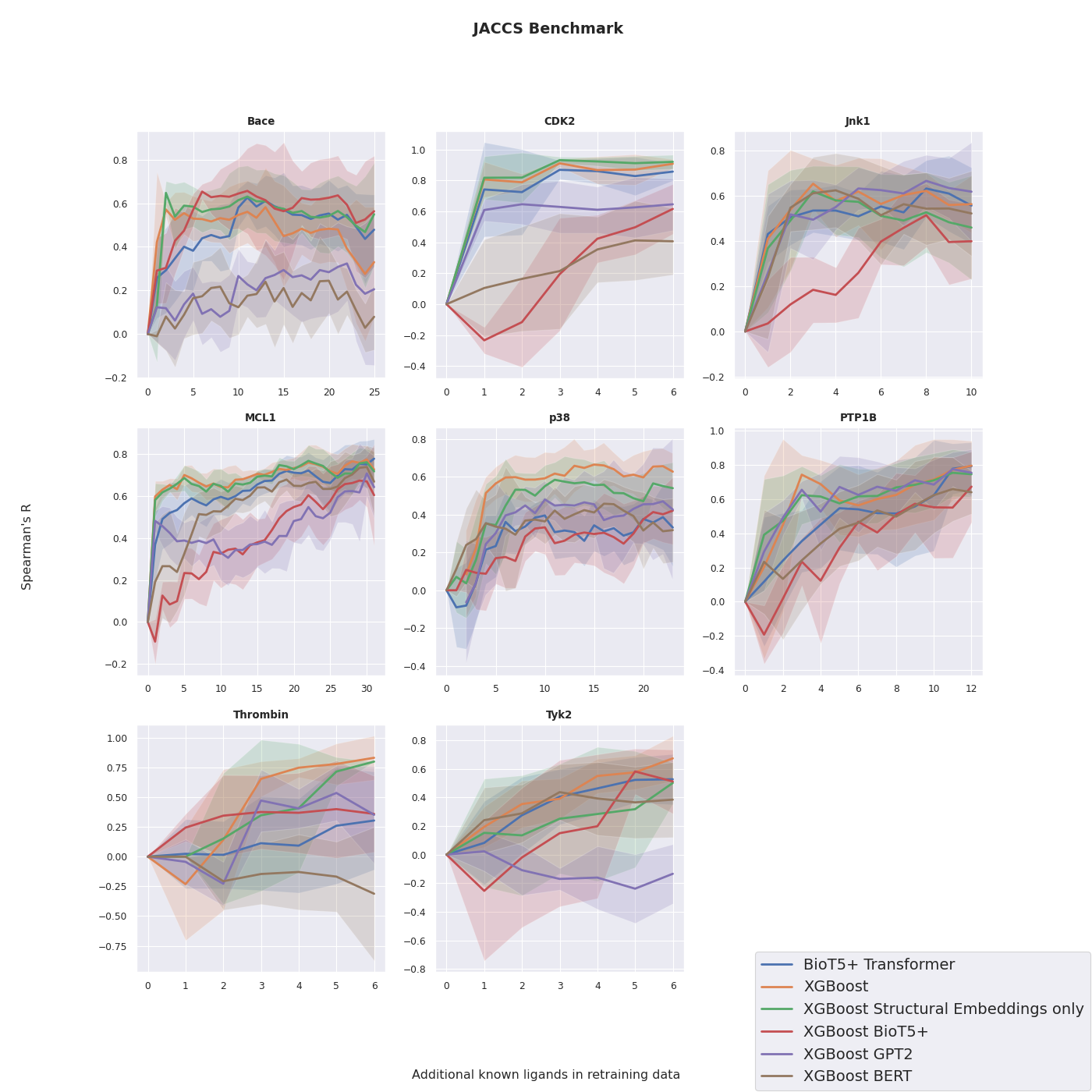}
      \caption{Additional Learning Curves on JACCS Benchmark Sets: This graph showcases a comparison of XGBoost models trained with RDKit embeddings against those using LLM embeddings from different architectures. Since LLM embeddings focus solely on structural information, the performance of an XGBoost model equipped only with structural RDKit embeddings is also depicted for comparison. Plotted is the Spearman's correlation against the pool of test molecules. The shaded area represents the standard deviation in the computed metric across the 25 model replicas.}
       \label{fig:schrodinger_bench_spear_extra}
\end{figure}

\subsection{Additional Plots Comparisons RDKit Embeddings on JACCS and Merck Sets}

\begin{figure}[H]
\centering
     \includegraphics[width=1.0\textwidth]{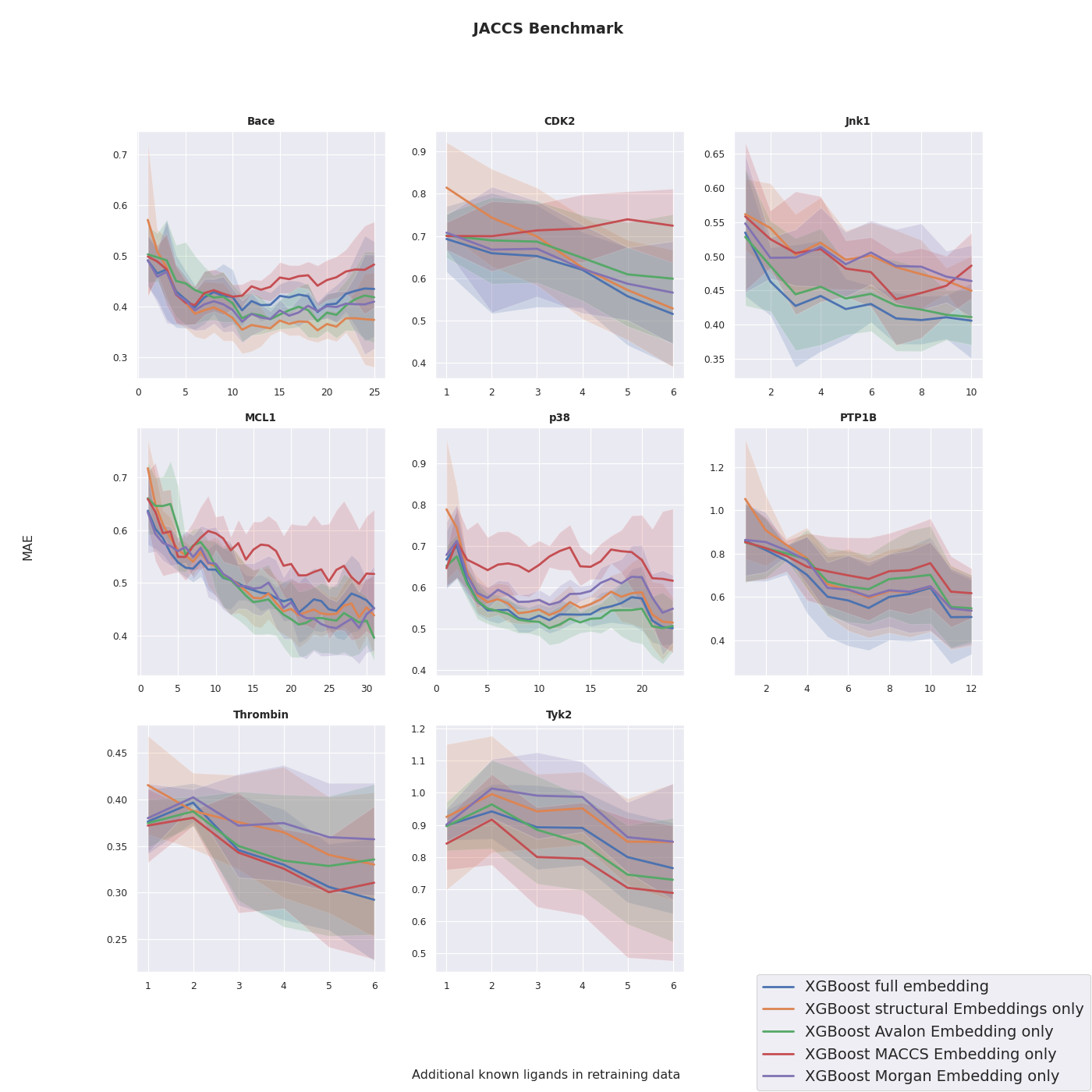}
      \caption{Learning Curves on the JACCS Benchmark Sets: This graph compares the performance of XGBoost models trained with combined and individual molecular RDKit embeddings. Plotted is the mean absolute error against the pool of test molecules. The shaded area represents the standard deviation in the computed metric across the 25 model replicas.}
       \label{fig:schrodinger_bench_pears_rdkit_comp}
\end{figure}

\begin{figure}[H]
\centering
     \includegraphics[width=1.0\textwidth]{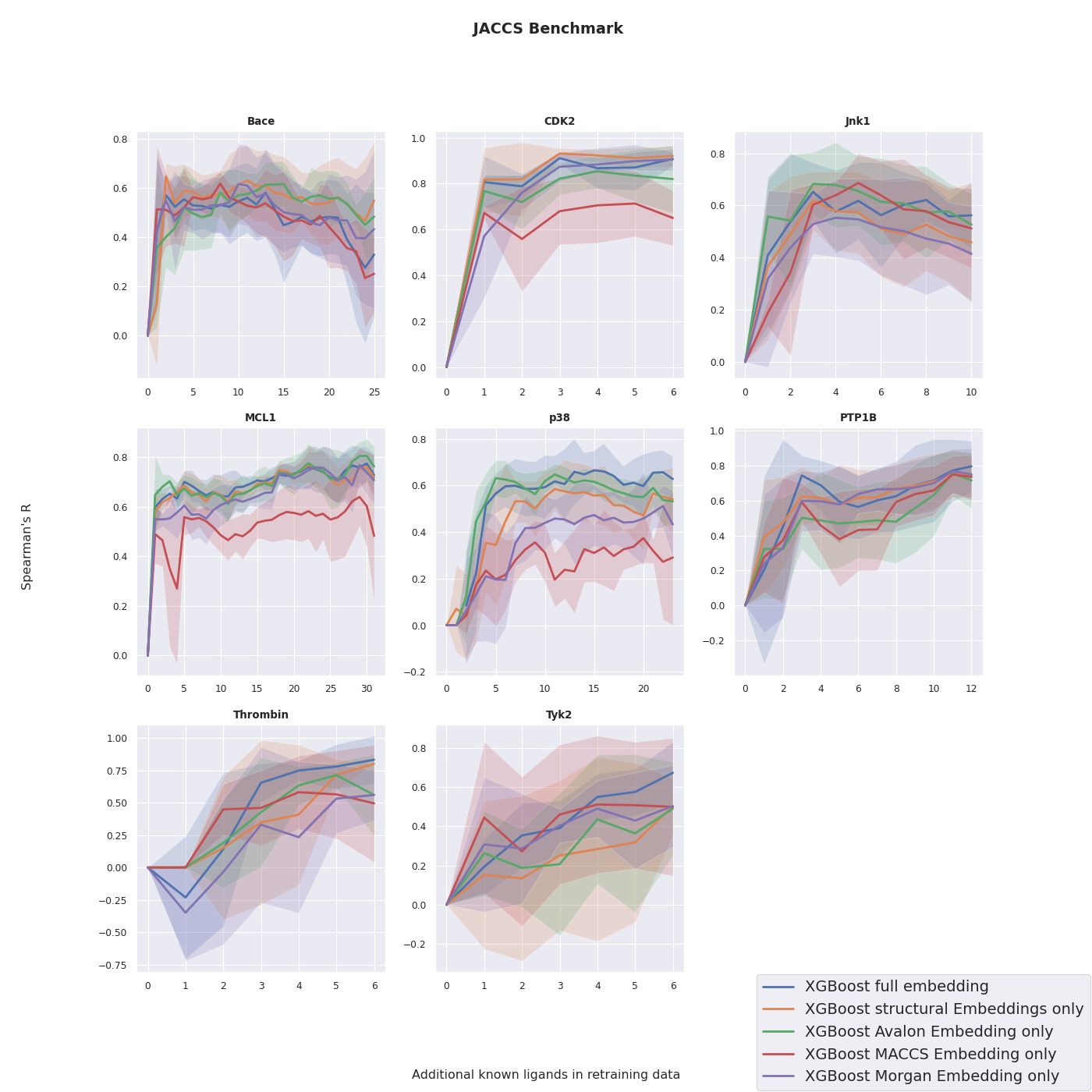}
      \caption{Learning Curves on JACCS Benchmark Sets: This figure displays the performance comparison of XGBoost models trained on combined and individual molecular RDKit embeddings. Plotted is the Spearman's correlation against the pool of test molecules. The shaded area represents the standard deviation in the computed metric across the 25 model replicas.}
       \label{fig:schrodinger_bench_pears_rdkit_comp}
\end{figure}

\begin{figure}[H]
\centering
     \includegraphics[width=1.0\textwidth]{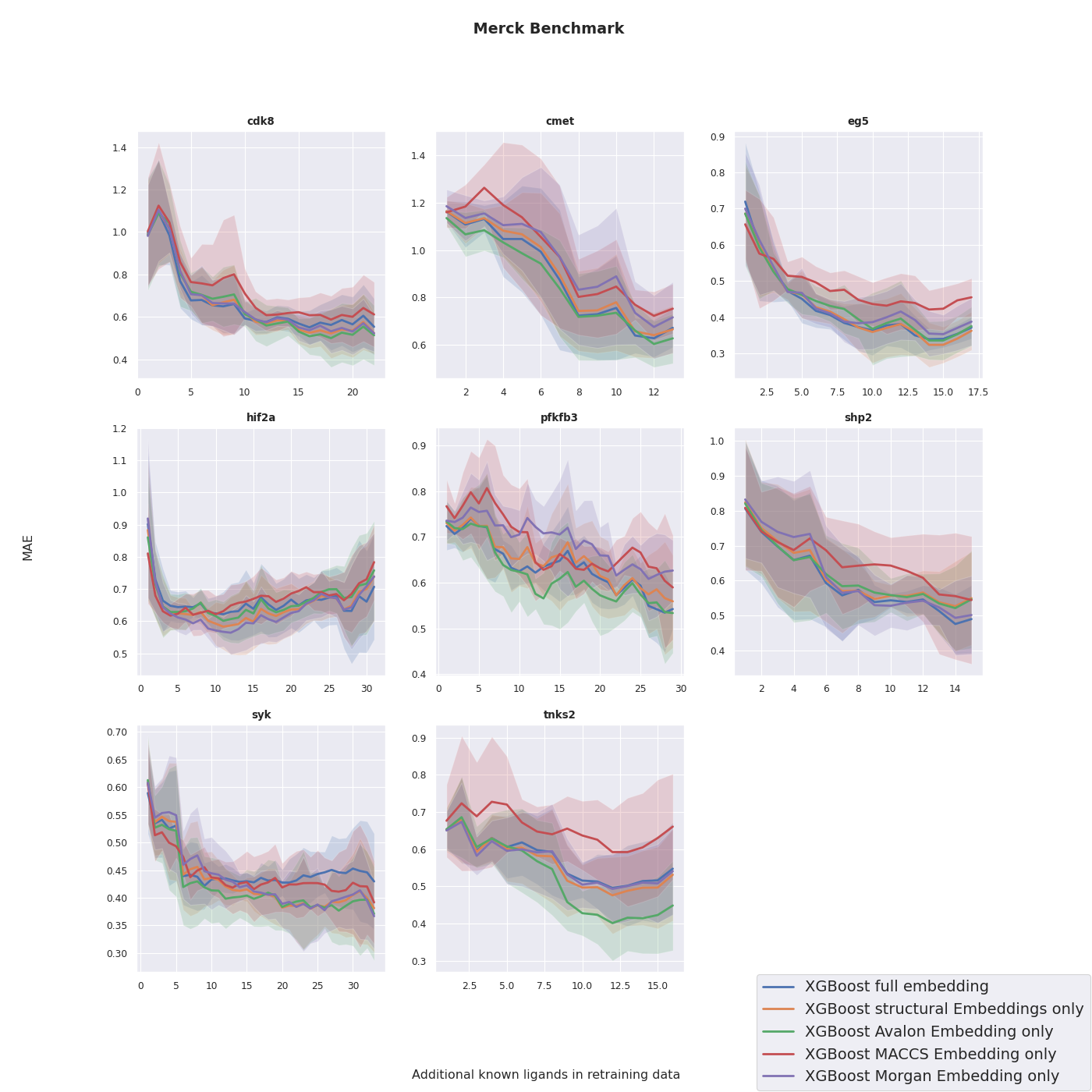}
      \caption{Learning Curves on Merck Benchmark Sets: This graph illustrates the performance comparison of XGBoost models trained with both combined and individual molecular RDKit embeddings. Plotted is the mean absolute error against the pool of test molecules. The shaded area represents the standard deviation in the computed metric across the 25 model replicas.}
       \label{fig:merck_bench_pears_rdkit_comp}
\end{figure}

\begin{figure}[H]
\centering
     \includegraphics[width=1.0\textwidth]{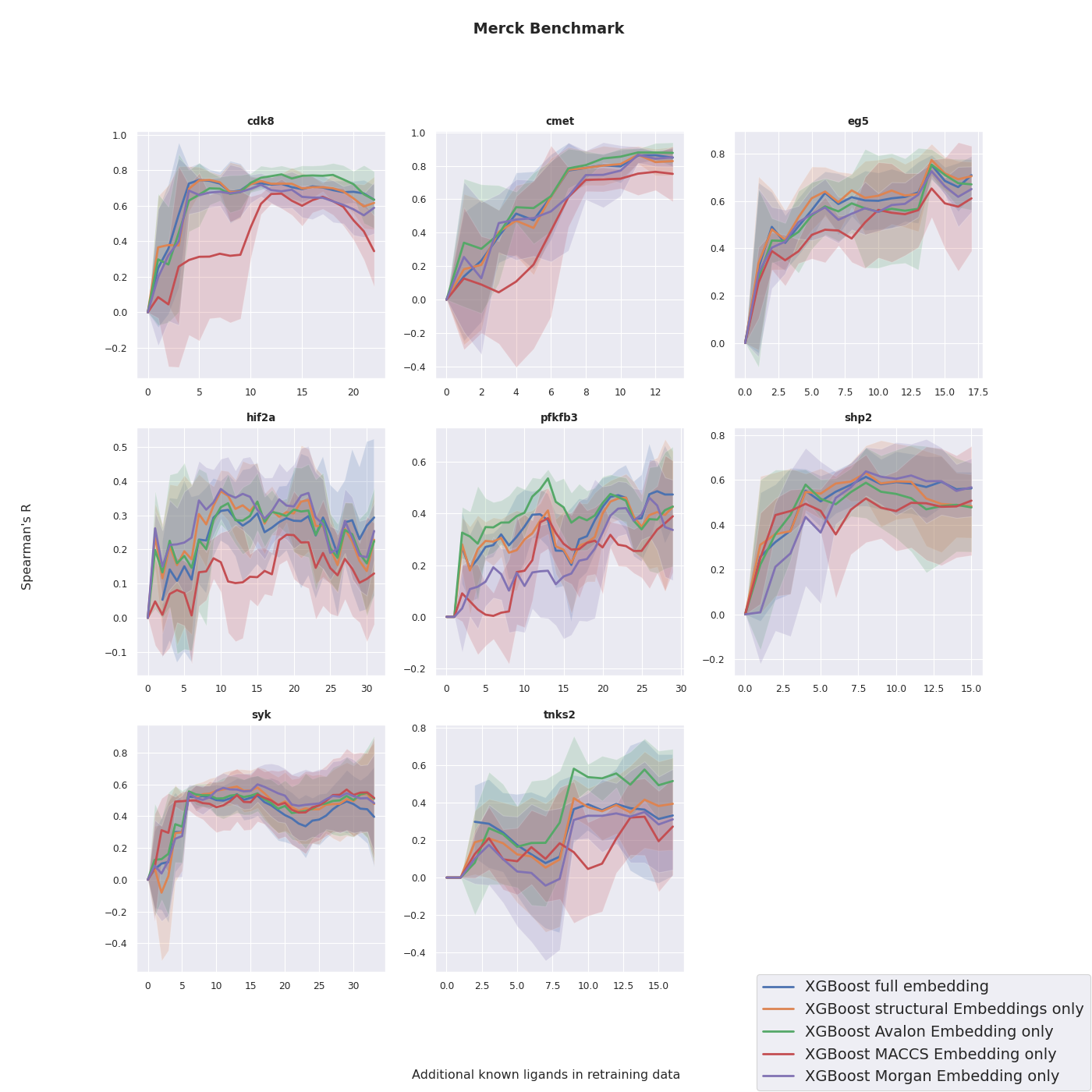}
      \caption{Learning Curves on Merck Benchmark Sets: This graph displays the performance of XGBoost models trained on both combined and individual molecular RDKit embeddings. Plotted is the Spearman's correlation against the pool of test molecules. The shaded area represents the standard deviation in the computed metric across the 25 model replicas.}
       \label{fig:merck_bench_pears_rdkit_comp}
\end{figure}

\subsection{Comparison of Number of Attention Heads in the Transformer Model on the Large Tyk2 Benchmark Set}

Transformer models are distinguished by their token-based attention mechanism, as initially described in \cite{transformer_architecture}. A critical hyperparameter in this architecture is the number of attention heads, which facilitates the parallelization of the attention process, allowing for a more comprehensive capture of relevant information at each layer. We further explore the impact of this hyperparameter in our study, specifically testing its effect in a transformer head model with 110K parameters, which is tailored for the downstream binding affinity prediction task.

\begin{figure}[H]
\centering
     \includegraphics[width=1.0\textwidth]{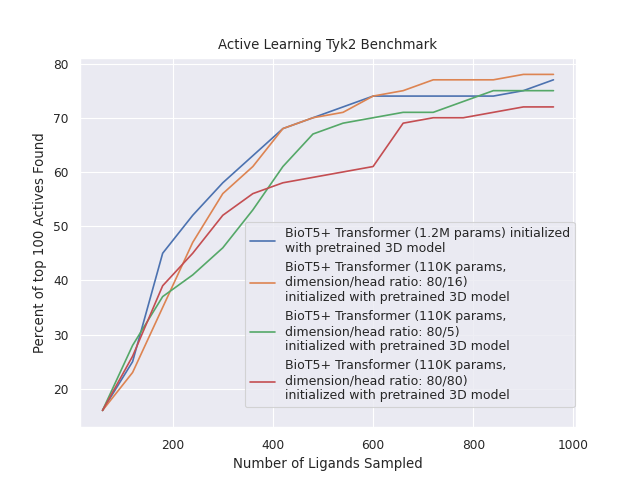}
      \caption{Learning Curve for the Tyk2 Benchmark Set: This graph displays the percentage of top 1\% binders found after each iteration of the active learning cycle. It compares the performance of a transformer model with fewer parameters, specifically focusing on the impact of varying the number of attention heads.}
       \label{fig:tyk2_heads_compare}
\end{figure}

From figure \ref{fig:tyk2_heads_compare}, it is evident that reducing the number of attention heads in the transformer model degrades its performance, as anticipated. Conversely, excessively increasing the number of heads, to the extent that per-token embedding vectors are reduced to very small sizes or even scalar values, also diminishes performance. This likely results from the loss of information when per-token vectors are excessively split between the heads. Therefore, a balance between the embedding dimension size and the number of heads is crucial to optimize performance.

\printbibliography